\documentclass[12pt]{article}
\usepackage{graphicx}
\usepackage{amsbsy}
\usepackage{amssymb}
\usepackage{bm}

\evensidemargin \oddsidemargin
\def\0{{\bf 0}}
\def \E {{\cal E}}
\def \LL {{\cal L}}

\def\b#1{{\mathbb #1}}

\newcommand{\bx}{{\bf x}}

\newcommand{\bxp}{{\bf x}^{\scriptscriptstyle \perp}}
\newcommand{\hbxp}{\hat {\bf x}^{\scriptscriptstyle \perp}}
\newcommand{\bXp}{{\bf X}^{\scriptscriptstyle \perp}}
\newcommand{\bX}{{\bf X}}
\newcommand{\bu}{{\bf u}}
\newcommand{\bup}{{\bf u}^{\scriptscriptstyle \perp}}
\newcommand{\hbup}{\hat {\bf u}^{\scriptscriptstyle \perp}}

\newcommand{\hbu}{\widehat{\bf u}}

\newcommand{\bwp}{{\bm w}^{\scriptscriptstyle \perp}}
\newcommand{\bv}{{\bf v}}

\newcommand{\rx}{{\rm x}}
\newcommand{\Ba}{{\bm \alpha}}
\newcommand{\Bap}{{\bm \alpha}^{\scriptscriptstyle \perp}}
\newcommand{\bb}{{\bm \beta}}

\newcommand{\Be}{{\bm \epsilon}}
\newcommand{\Bep}{{\bm \epsilon}^{\scriptscriptstyle \perp}}
\newcommand{\bE}{{\bf E}}
\newcommand{\bEp}{{\bf E}^{\scriptscriptstyle \perp}}
\newcommand{\hbEp}{\hat{\bf E}^{\scriptscriptstyle \perp}}

\newcommand{\bep}{{\bf e}^{\scriptscriptstyle \perp}}
\newcommand{\bbp}{{\bf b}^{\scriptscriptstyle \perp}}
\newcommand{\bB}{{\bf B}}
\newcommand{\bBp}{{\bf B}^{\scriptscriptstyle \perp}}
\newcommand{\hbBp}{\hat{\bf B}^{\scriptscriptstyle \perp}}

\newcommand{\EEpM}{E^{\scriptscriptstyle \perp}_{\scriptscriptstyle M}}
\newcommand{\bK}{{\bf K}}
\newcommand{\bKp}{{\bf K}^{\scriptscriptstyle \perp}}
\newcommand{\bA}{{\bf A}}
\newcommand{\bAp}{{\bf A}^{\scriptscriptstyle \perp}}

\newcommand{\bjp}{{\bf j}^{\scriptscriptstyle \perp}}

\newcommand{\Bp}{{\bf p}}
\newcommand{\bP}{{\bf P}}
\newcommand{\bPi}{{\bf \Pi}}

\newcommand{\bii}{\mathbf{i}}
\newcommand{\bj}{\mathbf{j}}
\newcommand{\bk}{\mathbf{k}}

\newcommand{\R}{\cal{R}}
\newcommand{\U}{{\cal U}}

\newcommand{\be}{\begin{equation}}
\newcommand{\ee}{\end{equation}}
\newcommand{\bea}{\begin{eqnarray}}
\newcommand{\eea}{\end{eqnarray}}
\newcommand{\ba}{\begin{array}}
\newcommand{\ea}{\end{array}}
\begin{document}
\title{Light-front approach to relativistic electrodynamics}

\author{  Gaetano Fiore$^{1,2}$   \\    
$^{1}$ Dip. di Matematica e Applicazioni, Universit\`a di Napoli ``Federico II'',\\
Complesso Universitario  M. S. Angelo, Via Cintia, 80126 Napoli, Italy\\         
$^{2}$       INFN, Sez. di Napoli, Complesso  MSA,  Via Cintia, 80126 Napoli, Italy}


\maketitle
\begin{abstract}
We illustrate how our recent light-front approach simplifies relativistic electrodynamics
with an electromagnetic  (EM) field $F^{\mu\nu}$ that is the sum of a (even very intense) plane travelling wave \ $F_t^{\mu\nu}(ct\!-\!z)$ \  and a static part \ $F_s^{\mu\nu}(x,y,z)$; it  adopts the light-like coordinate $\xi=ct\!-\!z$ instead of time $t$ as an independent variable. This can be applied to several cases 
of extreme acceleration,  both in vacuum and in a cold diluted plasma hit by a very short and intense laser pulse (slingshot effect, plasma wave-breaking and laser wake-field acceleration, etc.)
\end{abstract}

\section{Introduction}
\label{intro}

The equation of motion of a particle  of rest mass $m$, electric charge $q$  in an external EM  field  
\bea
\ba{l}
\displaystyle\dot\Bp(t)=q\bE[ct,\bx(t)] + \frac{\Bp(t) }{\sqrt{m^2c^2\!+\!\Bp^2(t)}}  \wedge q\bB[ct,\bx(t)] 
,\\[6pt]  
\displaystyle 
\dot \bx(t) =\frac{c\Bp(t) }{\sqrt{m^2c^2\!+\!\Bp^2(t)}}
\ea
\label{EOM}
\eea
in its  general  form is non-autonomous and highly nonlinear in the unknowns $\bx(t),\Bp(t)$. \
Here \ $\bx,\Bp$ \ are the  position
and   relativistic momentum of the particle, while \ $\bE,\bB$ are the electric and magnetic field;
we use  Gauss CGS units. 
In the situations characterizing many standard applications (\ref{EOM}) can be solved or simplified making one or more
of the following assumptions:

\begin{enumerate}

\item $\bE,\bB$ are constant or vary  ``slowly'' in space/time; 

\item the  motion of the particle keeps non-relativistic;  

\item $\bE,\bB$  are  so small  that nonlinear effects  in $\bE,\bB$ are negligible;  

\item $\bE,\bB$ are monochromatic waves, or very slow modulations thereof.

\end{enumerate}
If the role of the charges (isolated or in the form of a continuum) as sources of the EM field cannot be neglected, these equations are to be coupled with the Maxwell equations; the resulting system is much more complicated. 
In many interesting situations none of the above assumptions is satisfied.
For instance, many violent astrophysical electrodynamic processes
occur in the presence of extremely intense and rapidly 
varying electromagnetic fields (see e.g. \cite{TajNakMou17} and references therein).
On the other hand, the on-going, amazing developments of  laser technologies  leading to compact
sources of extremely intense and short coherent EM waves\footnote{The technique of {\it Chirped Pulse
Amplification}  \cite{StriMou85,MouTajBul06},  awarded with the Nobel Prize in physics in 2018, yields  pulses of intensity
up to $10^{23}$ Watt per square centimeter and duration down to  femtoseconds. Huge investments (e.g.  for the 
{\it Extreme Light Infrastructure}   program within the EU ESFRI roadmap) in  new technologies (thin film or  relativistic
mirror compression, etc.  \cite{MouMirKhaSer14,TajNakMou17}) 
will yield even more intense/short pulses, at a lower cost.}
allow extreme  accelerations through the Laser Wake Field Acceleration (LWFA)
mechanism \cite{Tajima-Dawson1979} in plasmas\footnote{In fact, huge investments aiming
at the construction of table-top particle accelerators based 
on Wake Field Acceleration are being made
all over the world. For instance, in Europe the large network of research centers ``European Plasma Research Accelerator with eXcellence 
In Applications" (EUPRAXIA) has been created to develop the associated technologies  and has just issued its expected {\it 
Conceptual Design Report} of reliable accelerators of this kind \cite{eupraxia1,eupraxia2,eupraxia3}. 
Among the present and future applications of accelerators we mention:

\begin{itemize}

\item  {\bf Medicine}: diagnostics (PET,...), cancer therapy by accelerated particles (electrons, protons, ions) or locally induced radioisotope production,...;

\item {\bf Research}: elementary particle physics, materials science, structural biology, inertial nuclear  fusion, electron beams for
X-ray free electron laser,...;

\item {\bf Industry}: atomic scale lithography, surface treatment of materials, sterilization,  additive-layer (i.e. 3D print)
manufacturing, detection systems across shields,...;

\item {\bf Environmental remediation}: flue gas cleanup, petroleum cracking, transmutation of nuclear wastes,....

\end{itemize} 
}.
To deal with such complex problems it is  is almost unavoidable to resort to numerical resolution methods, e.g. particle-in-cell (PIC)  techniques for plasmas. PIC or other codes in general involve huge and costly computations for each choice of the free parameters; exploring the parameter space blindly to pinpoint  the interesting regions is  prohibitive, if not accompanied by some analytical insight that can simplify
the work, at least in special cases or in a limited space-time region.
It is therefore important to look for new approaches simplifying the study of
the dynamics induced by  (\ref{EOM}) when conditions (i-iv)  are not fulfilled.

\begin{figure}
\includegraphics[width=10cm]{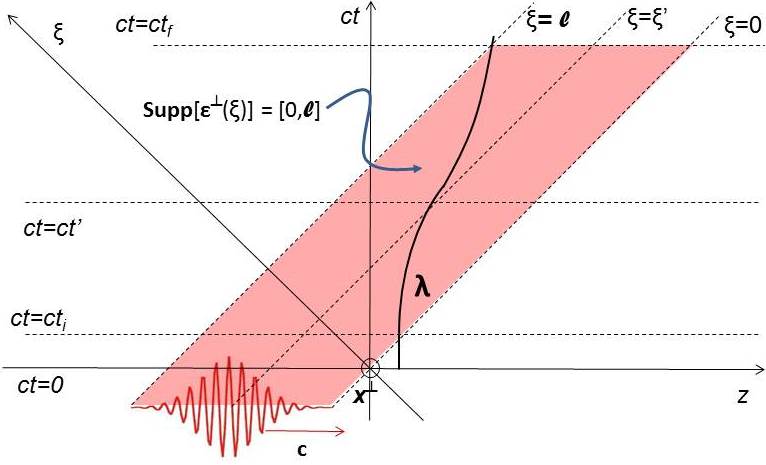}
\caption{Every worldline $\lambda$  and hyperplane $\xi\!=$const
in Minkowski space intersect once. The wave-particle interaction
occurs only along the intersection of $\lambda$ with the support of the EM wave (painted pink), which 
assuming (\ref{aa'}) is delimited by the  $\xi\!=\!0$ and $\xi\!=\!l$ hyperplanes.}
\label{Worldline}       
\end{figure}

\smallskip
Here we summarize an approach \cite{FioJPA18} that systematically
applies the light-front formalism \cite{Dir49}; it is especially fruitful if in the spacetime region of interest 
$\bE,\bB$  are the sum of
static parts and a plane transverse travelling wave  propagating in a fixed ($z$, say) direction:
\be
\bE(t,\bx)=\!\!\underbrace{\Bep(ct\!-\!z)}_{ travelling\, wave}\!+\underbrace{\bE_s(\bx)}_{static},\qquad \bB(t,\bx)=\underbrace{\bk\wedge\Bep(ct\!-\!z)}_{travelling\, wave} +
\underbrace{\bB_s(\bx)}_{static}.
\label{EBfields}
\ee
We decompose 3-vectors such as $\bx$ in the form $\bx\!=\!x\bii\!+\!y\bj\!+\!z\bk\!=\bxp\!+\!z\bk$ ($x,y,z$ are the  cartesian coordinates) and   denote as \ $\rx\!\equiv\!(ct,\bx)$ the set of coordinates of Minkowski spacetime points  with respect to the laboratory frame.  
We  assume only  that $\Bep(\xi)$ is piecewise continuous and
\bea
\ba{ll}
&\mbox{ {\bf a}) }\quad \Bep \mbox{ has a compact support }[0,l],\\[10pt] 
\mbox{ or} &\mbox{ {\bf a'})}\quad  \Bep \in L^1(\mathbb{R}),
 \ea   \label{aa'} 
\eea
and treat on the same footing all $\Bep$ fulfilling (\ref{aa'}) regardless of their Fourier analysis, in particular:
\begin{enumerate}
\item A modulated monochromatic wave (here $a_1^2\!+\!a_2^2\!=\!1$):
\be
\Bep\!(\xi)\!=\!\underbrace{\epsilon(\xi)}_{\mbox{modulation}}
\underbrace{[\bii a_1\cos (k\xi\!+\!\varphi)\!+\!\bj a_2\sin (k\xi)]}_{\mbox{carrier wave $\Be_o^{{\scriptscriptstyle \perp}}\!(\xi)$}}.
 \label{modulate}
\ee
\label{modula1}

\item 
A superposition of a finite number of waves of type 1.
\label{modula2}

\item An `impulse'  (few, one, or even a fraction of oscillation) \cite{Aki96,
MouMirKhaSer14}.

\end{enumerate}
The starting point of our approach is:  since no particle can 
reach the speed of light $c$, then $\tilde \xi(t)\!=\!ct\!-\!z(t)$ is strictly growing,
and  we can recast the eq. of motion (\ref{EOM}) in an equivalent 
 form ($\widehat{\ref{EOM}}$) making the change  of independent variable 
$t\mapsto \xi\!=\!ct\!-\!z$; the term $\Bep[ct\!-\!z(t)]$, where the unknown $z(t)$ is in the argument of the highly nonlinear and rapidly varying $\Bep$, becomes the known forcing term $\Bep(\xi)$, what makes 
(\ref{eqx}-\ref{equps0})
more manageable than  (\ref{EOM}). More radically, we can apply Hamilton's principle to the action functional $S(\lambda)$ of the particle parametrizing the worldline  
$\lambda$ (fig. \ref{Worldline}) by $\xi$ instead of $t$; the associated
Euler-Lagrange and Hamilton equations are equivalent and have the same features
as  (\ref{eqx}-\ref{equps0}). The Hamilton equations involve also a more convenient set of canonically conjugated momenta as unknown (dependent) variables.

We apply the approach first to an isolated particle (section \ref{GenRes}), then to a cold diluted plasma initially at rest and hit by a plane EM wave  (section \ref{Plasmas}).

\section{Electrodynamics of a single particle}
\label{GenRes}

\subsection{Set-up and general results}
Given points spacetime $\rx_0,\rx_1$  with $\rx_1$  in the causal cone of
 $\rx_0$, let $\Lambda$ be the set  of time-like curves  from $\rx_0$ to $\rx_1$. Given a $\lambda\in\Lambda$, we can equivalently use either $t$ or $\xi$ as a parameter on $\lambda$  in the corresponding  action functional of the particle: 
\bea
S(\lambda)=-\!\int_\lambda\! mc^2  d\tau+ qA(\rx)
=-\!\int\limits_{t_0}^{t_1}\!\!dt\,\underbrace{\frac{mc^2\!\!+ \!q  u^\mu A_\mu}{\gamma}}_{ L[\bx,\dot\bx,t]}
=-\!\int\limits_{\xi_0}^{\xi_1}\!\!\frac{d\xi}c\,\underbrace{\frac{mc^2\!\!+ \!q \hat u^\mu \hat A_\mu }{\hat s}}_{ \LL[\hat \bx,\hat \bx',\xi]}.    \label{Action}
\eea 
Here  \
$A(\rx)=A_\mu(\rx)d\rx^\mu=A^0(\rx)cdt\!-\!\bA(\rx)\cdot d\bx$ \ is the EM potential 1-form 
(the dot is the  scalar product in Euclidean $\b{R}^3$), which is related to the EM fields by \ $\bE=-\partial_t\bA/c-\nabla A^0$ 
and $\bB=\nabla\!\wedge\!\bA$   ($E^i\!=\!F^{i0}$, $B^1\!=\!F^{32}$, etc in terms of the EM tensor $F^{\mu\nu}$).   For any given function 
$f(ct,\bx)$ we  denote $\hat f(\xi, \hat \bx)\equiv f(\xi\!+\!  \hat z,  \hat \bx)$,
 abbreviate $\dot f\!\equiv\! df/dt$, $\hat f'\!\equiv\! d\hat f/d\xi$ (total derivatives). 
In particular, $\hat \bx(\xi)$ is the particle position seen as a function of $\xi$; 
it is determined  by $\hat \bx(\xi)=\bx(t)$. 
We raise and lower greek letter indices by the
Minkowski metric $\eta_{\mu\nu}\!=\!\eta^{\mu\nu}$, with
$\eta_{00}\!=\!1$, $\eta_{11}\!=\!-1$, etc.);  \ $(cd\tau)^2=(cdt)^2\!-\!d\bx^2$ 
is the square of the infinitesimal Minkowski distance ($\tau$ is the proper time of the particle), 
 $ dt/d\tau\!=\!\gamma\!=\!1/\sqrt{1\!-\! \bb^2}$ (with $\bb\equiv \dot\bx/c$)  is the Lorentz relativistic factor,
 $u\!=\!(u^0\!,\bu)
\!\equiv\!(\gamma,\!\gamma \bb)\!=\!\left(\!\frac {p^0}{mc^2},\!\frac {{\bf p}}{mc}\!\right)$  is the 4-velocity, i.e. the dimensionless version of the 4-momentum,  and 
\be
s\equiv\frac {d\xi}{d (c\tau)}=u^0\!- u^z=u^-=\gamma(1-\beta^z)>0          \label{defs0}
\ee
 is the light-like component $u^-$ of $u$, as well as the Doppler factor experienced by the particle, and is  positive-definite
[the first $=$ in (\ref{defs0}) follows from  $\gamma\!=\!dt/d\tau$, $p^z \!=\!m dz/d\tau$].  
We name $s$ the \ {\it light-like relativistic factor}, \ or shortly the  \  {\it $s$-factor}. \ In terms of the ``hatted" coordinates and their derivatives\footnote{In fact,
since \ $c\hat t(\xi)=\xi+\hat z$ \ implies \ $c\hat t'(\xi)=1+\hat z'(\xi)>0$, \ we find
\bea
\frac 1{\hat s}=\frac {d\tau}{dt}\frac{d(c t)}{d\xi}=
\frac 1{\gamma}\frac{d(c t)}{d\xi}=\sqrt{1\!-\! \left(\frac{d\bx} {cdt}\right)^2}\frac{d(c t)}{d\xi}=
\sqrt{ \left(\frac {cdt}{d\xi}\right)^2\!-\! \left(\frac{d\bx} {d\xi}\right)^2}=\sqrt{ \left(1\!+\!\hat z'\right)^2\!-\! \hat\bx'{}^2}
=\sqrt{1\!+\!2\hat z'\!-\!\hat\bx^{\scriptscriptstyle \perp}{}'{}^2}.               \nonumber    
\eea
}
\be
\hat\bu \!=\!\hat s \hat \bx',\qquad
\frac 1{\hat s}=
\sqrt{1\!+\!2\hat z'\!-\!\hat\bx^{\scriptscriptstyle \perp}{}'{}^2}.               \nonumber    
\ee
$\hat\gamma,\hat u^z,\hat\bb,\hat \bx'$ can be expressed as  
{\it rational  functions} of $\hat\bu^{{\scriptscriptstyle\perp}},\hat s$:
\bea
&&\hat\gamma\!=\!\frac {1\!+\!\hat\bu^{{\scriptscriptstyle\perp}}{}^2\!\!+\!\hat s^2}{2\hat s}, 
\quad \hat u^z\!=\!\hat\gamma\!-\!\hat s, 
 \quad \hat\bb\!=\! \frac{\hat\bu}{\hat\gamma},\quad \label{u_es_e} \\
 &&\hbxp{}'
=\displaystyle\frac {\hbup}{\hat s}, \qquad \quad  \hat z'
=\displaystyle\frac {1\!+\!\hbup{}^2}{2\hat s^2}\!-\!\frac 12 \label{eqx}
\eea
By Hamilton's principle, any extremum $\lambda$ of $S$  is the worldline of a possible motion of the particle with initial position $\bx_0$ at time $t_0$ and final position $\bx_1$ at time $t_1$. Hence it  fulfills Euler-Lagrange equations in both forms \
$\frac d {dt}\frac{\partial L}{\partial \dot \bx}=\frac{\partial L}{\partial \bx}$ and
$\frac d {d\xi}\frac{\partial \LL}{\partial \hat \bx'}=\frac{\partial \LL}{\partial \hat \bx}$,  \ equivalent to (\ref{EOM}).
The Legendre transforms  yield  the Hamiltonians $H\!\equiv\! \dot\bx\!\cdot\!\frac{\partial L}{\partial \dot \bx}\!-\!L\!=\!\gamma mc^2\!+\!q  A^0$ and  \ 
$\hat H\!\equiv\!\hat \bx'\!\cdot\!\frac{\partial \LL}{\partial \hat \bx'}\!-\!\LL\!=\!\hat\gamma mc^2\!+\!q\hat A^0$ \ respectively.
In fact, the change of  
`time' $t\mapsto\xi$ induces a {\it generalized canonical} (i.e. {\it contact}) transformation mapping $H\mapsto\hat H$. \
The latter is a {\it rational} function of $\hat \bx,\hat \bPi\!\equiv\!\frac{\partial \LL}{\partial \hat \bx'}$, or, equivalently, of $\hat s,\hat\bu^{{\scriptscriptstyle\perp}}$:
\bea
&&\hat H(\hat \bx,\!\hat\bPi;\!\xi)=mc^2\frac {1\!+\! \hat s^2\!\!+\! \hat \bu^{{\scriptscriptstyle\perp}2}\!}{2\hat s}\! +\! q\hat A^0\!(\xi,\hat\bx),\qquad
\label{Ham} \\ &&
\mbox{where}\:\: 
 \left\{ \!\!
\ba{l} \displaystyle mc^2\hat\bu^{{\scriptscriptstyle\perp}}\!\!=\!\hat\bPi^{{\scriptscriptstyle\perp}}\!
\!-\!q\hat \bA^{{\scriptscriptstyle\perp}}(\xi,\hat\bx),\\[6pt]
\displaystyle mc^2\hat s\!=\!-\hat \Pi^z\!\!-\!q[\hat A^0\!-\!\hat A^z](\xi,\hat\bx),        
\ea \right.     
\qquad
\eea
 while \ $H(\bx,\bP,t)\!=\!\sqrt{\!m^2c^4\!+\!(c\bP\!-\!q\bA)^2}+\!qA^0$ 
\ (where $\bP\!\equiv\!\frac{\partial L}{\partial \dot \bx}\!=\!\Bp\!+\!\frac qc\bA$) is not.
Eq. (\ref{EOM})  are also equivalent to the Hamilton equations \
$\hat \bx'\!=\!\frac{\partial \hat H}{ \partial \hat\bPi},\: \hat\bPi' \!=\!-\frac{\partial \hat H}{ \partial \hat \bx}$. These amount to (\ref{eqx}) and 
\bea
\ba{l}
 \displaystyle\hat\bup{}' =\frac q{mc^2\, \hat s}\!\left[\hat\gamma\hat\bE\!+\!\hat\bu\!\wedge\!\hat\bB\right]^{\scriptscriptstyle \perp}\!
, \\[12pt]
 \displaystyle \hat s'  = \frac {q}{mc^2}\!\left[\frac{\hat\bup}{\hat s}\!\cdot\!\hat\bEp\!-\!\hat E^z\!-\!\frac{(\hat\bup\!\wedge\!\hat\bBp)^z}{\hat s}\right]
\ea
\label{equps0}
\eea
with $\hat \gamma$ as given in (\ref{u_es_e}). 
All the new equations, in particular these ones, can be also obtained more  directly from the old ones
by putting a caret on all dynamical variables and replacing \ $d/dt$ by $(c\hat s/\hat \gamma)d/d\xi$.
Along the solutions  $\hat H$ gives the particle energy  as a function of $\xi$.
Once solved (\ref{eqx}-\ref{equps0}), analytically or numerically, 
we just need to invert $\hat t (\xi)\!=\!\xi\!+\!\hat z(\xi)$ and set 
$\bx(t)=\hat\bx[\xi(t)]$,... to obtain the solution as a function of $t$.

\medskip
Under the EM field  (\ref{EBfields}) eqs  (\ref{equps0}) amount to 
\bea
\ba{l}\displaystyle\hbup{}'\!=\frac q{mc^2}\!\left[(1\!+\!\hat z')\hbEp_s\!+\!(\hat\bx'\!\wedge\!\hat\bB_s)^{\scriptscriptstyle \perp}\!+\!\Bep(\xi) \right]\!,\\[10pt]
\displaystyle\hat s'=\frac {-q}{mc^2}\left[\hat E^z_s-\hbxp{}'\!\cdot\!\hbEp_s\!+(\hbxp{}'\!\wedge\!\hbBp_s)^z\right]\!,    
\ea\qquad \label{equps} 
\eea
while the  {\it energy gain}  (normalized to $mc^2$) in the interval $[\xi_0,\xi_1]$ is 
\be
\E\equiv \frac{\hat H(\xi_1)\!-\!\hat H(\xi_0)}{mc^2}= \int^{\xi_1}_{\xi_0}\!\!\! d\xi\,\frac{q\Bep}{mc^2}\!\cdot\!\frac{\hbup}{\hat s }.
\label{EnergyGain}
\ee 
We can obtain the fields (\ref{EBfields})  from an EM potential of the same form, \
$A^\mu(\rx)=\alpha^\mu(ct\!-\!z)+A_s^\mu(\bx)$; \
in the Landau gauges ($\partial_\mu A^\mu\!=\!0$) \ $\bA_s$ must fufill the
Coulomb gauges  ($\nabla\!\cdot\! \bA_s\!=\!0$), and it must be \ $\alpha^z{}'=\alpha^0{}'$, 
 \ $\Bep\!=\!-\Bap{}'$, $\bE_s\!=\!-\nabla\! A_s^0$, $\bB_s\!=\!\nabla \!\wedge\! \bA_s$. \
We shall set  $\alpha^z=\alpha^0=0$, as $\alpha^z,\alpha^0$  appear neither in the observables $\bE,\bB$ 
\ nor in the equations of motion.
We fix $\Bap(\xi)$ uniquely by requiring that
it vanish as $\xi\to-\infty$; this leads to
\bea
\Bap(\xi)\!\equiv\!-\!\!\int^{\xi}_{ -\infty }\!\!\!\!\!\!\!dy\,\Bep\!(y).         \label{defBap}
\eea
Under rather general assumptions on $\epsilon$ (\ref{modulate}) implies
\be
\Bap(\xi)= -   \frac {\epsilon(\xi)}k \:\Bep_p\!(\xi)+O\left(\frac 1 {k^2}\right)
 \simeq -   \frac {\epsilon(\xi)}k \,\Bep_p\!(\xi)              \label{slowmodappr}
\ee
where $\Bep_p\!(\xi)\! :=\!\Bep_o(\xi\!+\!\pi/2k)$; in the appendix of \cite{FioJPA18} we  recall upper bounds  for the remainder $O(1/k^2)$.
For very slow modulations  (i.e. $|\epsilon'|\!\ll\! |k\epsilon|$, namely the modulating amplitude $\epsilon$ does not vary significantly over the wavelength $\lambda\!\equiv\!2\pi/k$) 
 - like the ones characterizing most conventional applications (radio broadcasting, ordinary laser beams,  etc.)  - the right estimate is  very good.
Consequently, if $\epsilon(\xi)$ goes to zero also as $\xi\!\to\!\infty$, then 
$\Ba\!^{{\scriptscriptstyle\perp}}(\xi)$ approximately does  as well.
In particular,  in case {\bf a}) eq. (\ref{defBap}) implies
$\Bap(\xi)\!=\!0$ if $\xi\!\le\! 0$,  $\Bap(\xi)\!=\!\Bap(l)\simeq 0$ if $\xi\!\ge\! l$. \

While the usual Hamilton equations 
$\dot\bP =-\partial  H/ \partial \bx$, $\dot\bx=\partial H/ \partial \bP$,  which amount to
\bea
\dot\bu(t)=\frac q{mc}\!\left\{\bE_s\!\!+\!\left(\!\frac{\dot\bx}c\!\wedge\!\bB_s\!\right)\!\!+\! \left(\!\frac{\dot\bxp}c\cdot \Bep[ct\!-\!z(t)]\!\right)\!\bk
\!-\!\frac 1c \frac d{dt}\Bap[ct\!-\!z(t)]\right\}\!, \label{equ}
\eea
and $\dot\bx=\bu/\sqrt{1\!+\!\bu^2}$, have no rational form  and contain the unknown $ct\!-\!z(t)$  in the argument of the rapidly varying function $\Bep,\Bap$,  (\ref{eqx}-\ref{equps}) on the contrary are rational
in the unknowns and contain $\Bep(\xi)$ as a known forcing term.
This simplifies their study and the determination
of $\E$.

\subsection{\small Dynamics under an EM potential of the form \  $A^\mu\!=\!A^\mu(t,\!z)$}
\label{A=A(t,z)}

\begin{figure}[h]
\begin{minipage}{.47\textwidth}
\includegraphics[width=7.5cm]{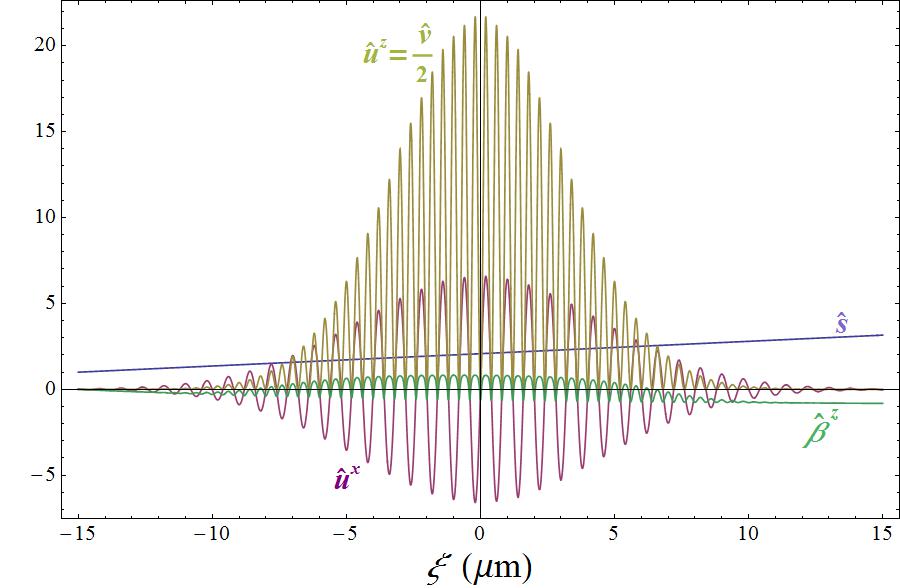}
\\ \includegraphics[width=7.5cm]{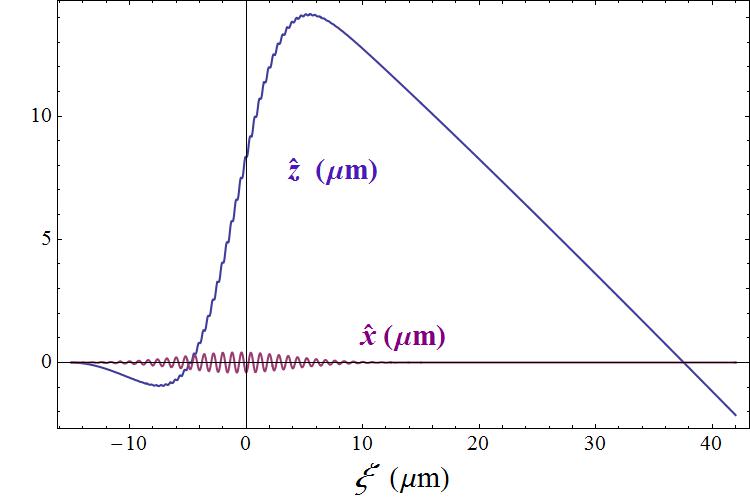}
\end{minipage}%
\hfill\begin{minipage}{.5\textwidth}
\includegraphics[width=8cm]{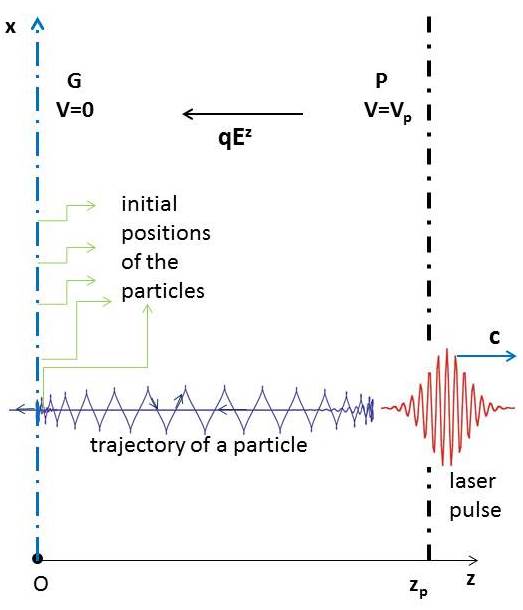}
\end{minipage}
\caption{Left: the motion (\ref{Ezcost})   induced by a linearly polarized modulated EM wave  (\ref{modulate}) with wavelength $\lambda\!=\!2\pi/k\!=\!0.8\mu$m, gaussian enveloping amplitude $\epsilon(\xi)\!=\! \EEpM\exp[-\xi^2/2\sigma]$ with
$ \sigma \!=\! 20\mu$m$^2$ 
and $|q|\EEpM/kmc^2\!=\!6.6$,  trivial initial conditions, $\bB_s \!=\!\0$,
$\bE_s \!=\!\bk E^z_{\scriptscriptstyle M}$, where $E^z_{\scriptscriptstyle M}q\!\simeq\!  37$GeV/m
(with such a wave this yields the maximum energy gain, $\E_f\!\simeq\! 1.5$). Right: the corresponding trajectory in the $zx$ plane within an hypothetical acceleration device based on a laser pulse and 
metallic gratings $G,P$ at potentials $V\!=\!0,V_p$, with $qV_p/z_p\!\simeq\! 37$GeV/m. 
}
\label{Ez=const>}
\end{figure}

Eq. (\ref{equps0})  are further simplified if  \  $A^\mu$ does not depend on transverse coordinates.  \ \ 
This applies in particular  if $\bE_s= E_s^z(z)\bk$, $\bB_s=\bB_s^{\scriptscriptstyle \perp}(z)$, choosing \ $A^0=-\int^z\!d\zeta E_s^z(\zeta)$, 
\ $\bAp\!=\Bap\!-\!\bk\wedge\!\int^z\!d\zeta \bBp(\zeta)$, $A^z\!\equiv\! 0$.
As $\partial \hat H/\partial \hat \bx^{\scriptscriptstyle \perp}\!=\!0$, we find
$\hat\bPi^{\scriptscriptstyle \perp}\!=\!q\bKp\!=\!\mbox{const}$, i.e.
the known result $\frac {mc^2}q\hat\bu^{\scriptscriptstyle \perp}\!=\!\bKp
\!-\!\hat \bA^{{\!\scriptscriptstyle\perp}}\!(\xi,\!\hat z)$.
  Setting $v\!:=\!\hbup{}^2$  and replacing  in   (\ref{equps}) we obtain 
the first order system in the unknowns $\hat z,\hat s$
\bea
\hat z'=\displaystyle\frac {1\!+\!\hat v}{2\hat s^2}\!-\!\frac 12, \qquad
\hat s'=\frac{-q}{mc^2}E_s^z(\hat z)\!-\!\frac 1{2\hat s}
\frac{\partial \hat v}{\partial \hat z}.     \label{reduced}
\eea
Once solved system (\ref{reduced}) for $\hat z(\xi), \hat s(\xi)$,
the other unknowns are  obtained integrating (\ref{eqx}):
\bea
 \hat\bx(\xi)=\bx_0+\!\displaystyle\int^\xi_{\xi_0}\!\!\! dy \,\frac{\hat\bu(y)}{\hat s(y)}  \label{hatsol}
\eea
\smallskip
[the $z$-component of (\ref{hatsol}) amounts to  (\ref{reduced}a)  with initial condition $\hat z(\xi_0)\!=\!z_0$]. 
If in addition $\bB_s\!\equiv\!0$, then 
 $\bA_s\!\equiv\!0$  (in the Coulomb gauge), implying that
\ $\hat\bu^{\scriptscriptstyle \perp}\!(\xi)\!=\!\frac q{mc^2}\left[\bK^{\scriptscriptstyle \perp}\!-\Bap(\xi)\right]$ and 
$ \hat v\!=\!\hat\bu^{{\scriptscriptstyle\perp}2}$ in   (\ref{reduced})
 are already known. Thus the system   (\ref{reduced}) to  be solved simplifies to
\bea
&& \hat z'=\frac {1\!+\! \hat v}{2\hat s^2}\!-\!\frac 12, \qquad  \hat s'=\frac{-q}{mc^2}E_s^z(\hat z).
 \label{heq1r} 
\eea
{\bf Remarks.} Some noteworthy properties of the corresponding solutions are  \cite{FioJPA18}:  

\begin{enumerate}

\item Where $\Bep(\xi)\!=\! 0$ then $ \hat v(\xi)\!=\!v_c\!=$const,
$\hat H$ is conserved,  (\ref{heq1r}) is solved by quadrature.

\item 
\label{notransv}
In case (\ref{aa'}a) the final transverse momentum is \ $mc\hat\bup(l)$.
 If  $\epsilon$ of (\ref{modulate}) varies very slowly and $\hat\bup\!(0)\!=\!\0$, then by (\ref{slowmodappr}) 
$\hat\bup(l)\!\simeq\!0$.

\item 
Fast  oscillations of $\Bep$ make $\hat z(\xi)$ oscillate much less than $\hat\bxp(\xi)$, and 
$\hat s(\xi)$ even less: as $\hat s\!>\!0$, $\hat v\!=\!\hbup{}^2\!\ge\! 0$, 
integrating (\ref{heq1r}a) averages the fast oscillations of $\bup$ to yield much smaller relative 
oscillations of $\hat z$,  while integrating (\ref{heq1r}b) averages the residual small 
oscillations of $E_s^z[\hat z(\xi)]$ to yield an essentially smooth $\hat s(\xi)$.
On the contrary, $\hat \gamma(\xi), \hat\bb(\xi), \hat\bu(\xi),...$, which 
are recovered via (\ref{u_es_e}), oscillate fast, and so do also $\gamma(t), \bb(t), \bu(t),..$. See e.g. fig. \ref{Ez=0},\ref{Ez=const>},\ref{graphs}.
\label{insensitive}

\item If  $\bup\!(0)\!=\!\0$ and the EM wave is a very slowly modulated (\ref{modulate})-(\ref{aa'}a),
integrating   (\ref{EnergyGain}) by parts across $[0,l]$ and using  (\ref{slowmodappr}) we find that the final energy gain is given by
\be
\E_f=\int^{l}_{0}\! d\xi \: \frac {\hat  v'(\xi)}{2\hat s(\xi)}\simeq
\int^{l}_{0}\! d\xi \: \frac {\hat  v(\xi)\hat s'(\xi)}{2\hat s^2(\xi)}; \label{DeltaHf}
\ee 
this   will be automatically positive (resp. negative) if $\hat s(\xi)$ is growing 
(resp. decreasing) in all of $[0,l]$. Correspondingly, the  interaction with the EM wave can
be used to accelerate (resp. decelerate) the particle. 
\label{signEnGain}

\end{enumerate}

\subsection{Dynamics under travelling waves and static fields $\bE_s,\bB_s$}

If $\bE_s,\bB_s\!=$const
then 
eq. (\ref{equps}) are immediately integrated to yield 
\bea
\ba{l}
\hat\bup=\displaystyle\frac q{mc^2}\left[\bKp\!-\Bap(\xi)\!+(\xi\!+\!\hat z)\bEp_s+
(\hat\bx\!\wedge\!\bB_s)^{\scriptscriptstyle \perp}\!\right],\\[12pt]
\hat s=\displaystyle\frac {-q}{mc^2}\left[K^z\!+\xi E^z_s-\hat \bxp \!\cdot\!\bEp_s\!+(\hat\bxp\!\wedge\!\bB_s)^z\right]
\ea    \label{constEsBs}
\eea
(the integration constants $K^j$ are fixed by the initial conditions), or more explictly
\bea
\ba{l}
\hat u^x= w^x(\xi)\!+\! (e^x\!-\!b^y)\hat z\!+\!b \hat y,\\[10pt]
\hat u^y= w^y(\xi)\!+\! (e^y\!+\!b^x)\hat z\!-\!b \hat x,\\[10pt]
\hat s=w^z(\xi)\!+\!(e^x\!-\!b^y) \hat x \!+\!(e^y\!+\!b^x)\hat y;
\ea    \label{constEsBs'}
\eea
here we have introduced the dimensionless functions \ $\bwp(\xi)\! :=\! q\left[\bKp\!\!-\!\Bap(\xi)\!+\xi\bEp_s\right]/mc^2$, $w^z(\xi)\! :=\!-q(K^z\!\!+\!\xi E^z_s)/mc^2$ \ and the constants \ $\bep:=\! q\bEp_s/mc^2$, \ $\bbp\!+\!b\bk\! :=\! q\bB_s/mc^2$. 
Hence,
it remains to solve the system of three first order  equations in rational form  in the unknowns $\hat x,\hat y, \hat z$ that is obtained
replacing (\ref{constEsBs'}) in (\ref{eqx}).

\medskip
\noindent
{\bf Proposition 9 in \cite{FioJPA18}.} \ 
{\it If  $\bE_s,\bB_s$ are constant fulfilling the only condition \ 
$\bBp_s\!=\!\bk\wedge\bEp_s$ then, setting $\kappa\!:=\!qE_s^z/mc^2$,
$x_0\!:=\!\hat x(\xi_0)$, $y_0\!:=\!\hat y(\xi_0)$, $z_0\!:=\!\hat z(\xi_0)$, 
 $s_0\!:=\!\hat s(\xi_0)$, we can put the solutions of (\ref{eqx}) in the compact form}
\bea
\ba{l}
\displaystyle    
( \hat x +i \hat y)(\xi)=(s_0\!-\!\kappa\xi)^{ib/\kappa}\left[ \frac{( x_0 +i y_0)}
{(s_0\!-\!\kappa\xi_0)^{ib/\kappa}}+
\int^\xi_{\xi_0}\!\!\!\!  d\zeta \, \frac{(w^x+iw^y) (\zeta)}
{(s_0\!-\!\kappa\zeta)^{1+ib/\kappa}} \right],\\[12pt]
 \displaystyle  \hat s(\xi)=s_0\!-\!\kappa\xi, \qquad  \hat u^z(\xi)\!=\!\frac {1}{2(s_0\!-\!\kappa\xi)}+(s_0\!-\!\kappa\xi)\,\frac {\hat \bxp{}'{}^2(\xi)-1}{2},\\[12pt]
\displaystyle \hat\bup\!(\xi)\!=\!(s_0\!-\!\kappa\xi)\,\hat \bxp{}'(\xi),   \qquad    \quad     \hat \gamma(\xi)\!=s_0\!-\!\kappa\xi\!+\!\hat u^z(\xi),\\[12pt]
\displaystyle\hat z(\xi)\!=\!z_0+\int^\xi_{\xi_0} \frac {d\zeta }{2}
\left[\frac {1}{(s_0\!-\!\kappa\zeta)^2}\!+\!\hat \bxp{}'{}^2(\zeta)\!-\!1\right].
\ea   \label{SolEqBzEzg}
\eea
Up to our knowledge such general solutions 
have not appeared  in the literature before Ref. \cite{FioJPA18}, but reduce to known ones under additional assumptions on $\bE_s,\bB_s$. Now we briefly review some of them, adopting for simplicity the initial conditions $\bx_0\!=\!\0\!=\!\bu_0$, whence $s_0=1$.

\subsubsection{Zero static fields case: $\bE_s\!=\!\bB_s\!=\!\0$.}

Then (\ref{SolEqBzEzg}) becomes \cite{LanLif62,Fio14JPA}:
\bea
\ba{l}
\displaystyle\hat s\!\equiv\! 1, \quad  \hat\bu^{\scriptscriptstyle \perp}\!\!=\!
\frac {-q\Bap}{mc^2}, \quad \hat u^z\!=\!\frac {\hat  \bup{}^2}2, \quad\hat\gamma\!=\!1\!+\!\hat u^z\\[16pt] 
\displaystyle \hat z(\xi)\!=\! \int ^\xi_{\xi_0}\!\!\!\!dy\,\frac{\hat\bu^{{\scriptscriptstyle\perp}2}(y)}2,\quad \hbxp\!(\xi)\!=\! \int ^\xi_{\xi_0}\!\!\!\!dy\,\hbup\!(y).      \ea                      \label{U=0s=0}
\eea
\begin{figure}[h]
\centering
\includegraphics[width=7.9cm]{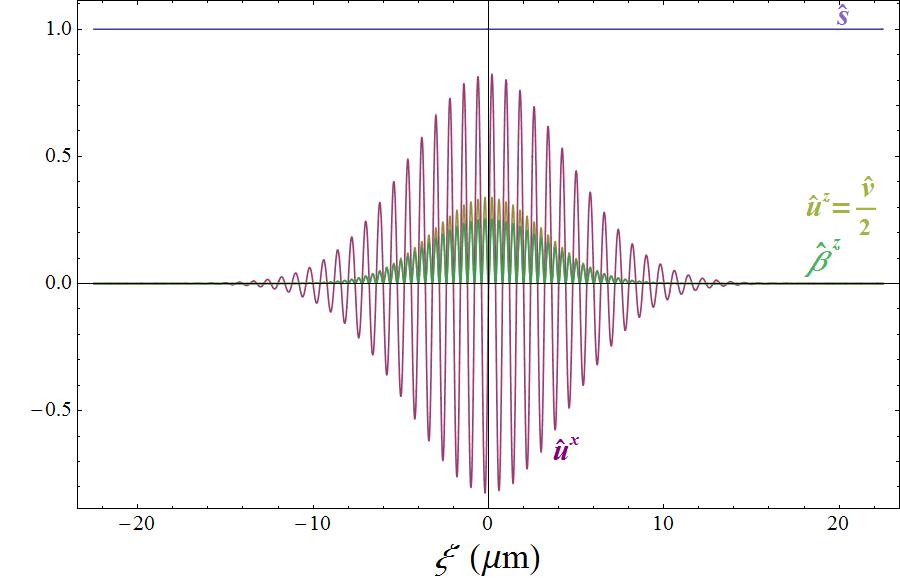} \hfill  \includegraphics[width=7.9cm]{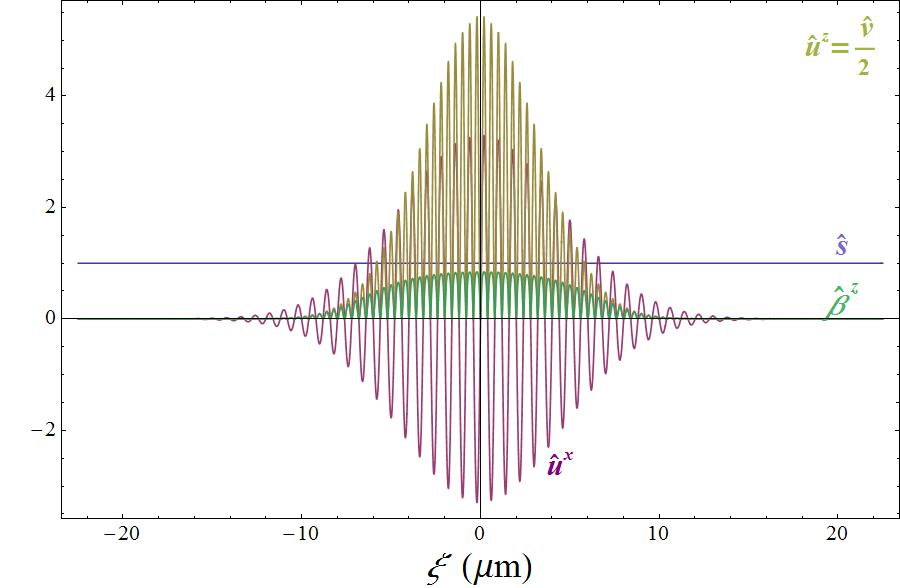}\\
\includegraphics[width=7.9cm]{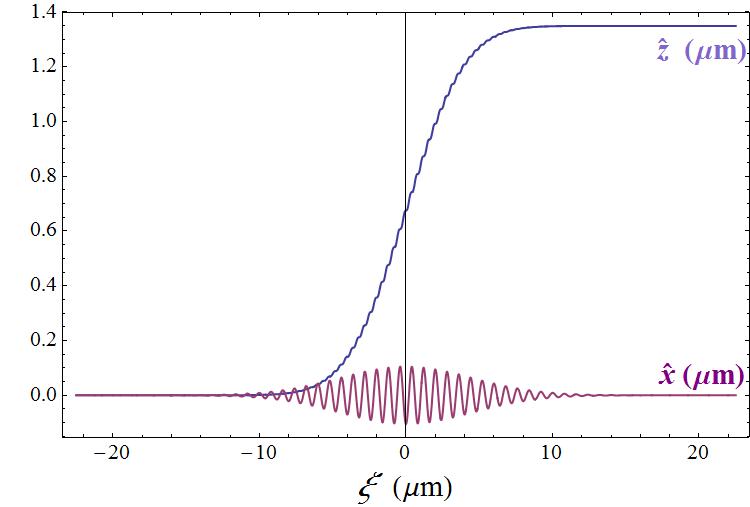}\hfill \includegraphics[width=7.9cm]{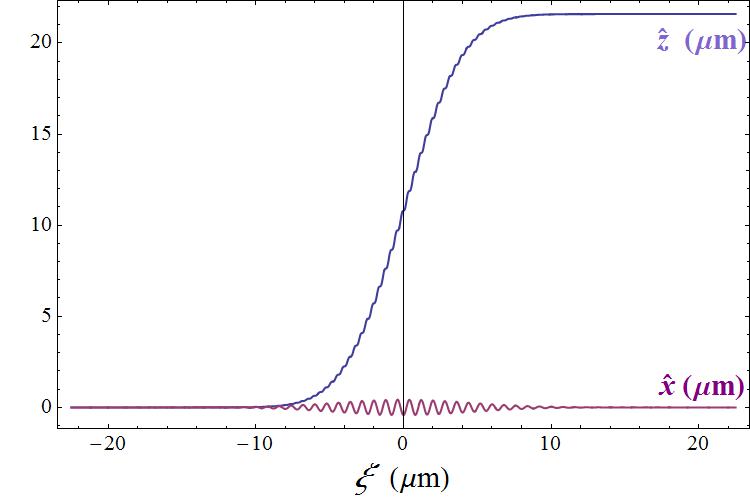} \\
\vskip.3cm
\includegraphics[width=5.7cm]{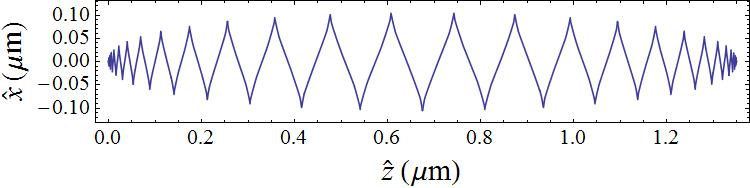} \hfill 
\includegraphics[width=10.1cm]{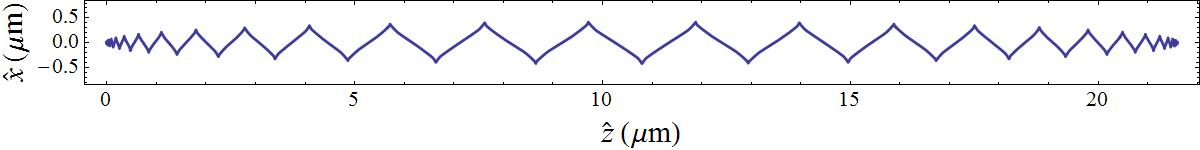}
\caption{Solutions (\ref{U=0s=0}) and $e^-$ trajectories  in the $zx$ plane induced 
by two $x$-polarized pulses with 
carrier wavelength $\lambda\!\equiv\!2\pi/k\!=\!0.8\mu$m, gaussian modulation $\epsilon(\xi)=\EEpM\exp[-\xi^2/2\sigma]$,
$ \sigma \!=\! 20\mu$m$^2$, \ $a_0\!\equiv\! e \EEpM/kmc^2  \!=\! 0.8$ (left) or 
$a_0 \!=\!  3.3$ (right). 
}
\label{Ez=0}      
\end{figure}
The solutions (\ref{U=0s=0}) induced by two $x$-polarized pulses 
and the corresponding $e^-$ trajectories  in the $zx$ plane  are shown in fig. \ref{Ez=0}. Note that:
\begin{itemize}

\item The maxima of $u^z$, 
$\Bap{}^2$  (and approximately also of $\epsilon(\xi)$, if $\epsilon(\xi)$ is slowly varying) coincide.

\item  Since $u^z\!\ge\!0$,  the $z$-drift is positive-definite.
Rescaling $\Bep\mapsto a\Bep$, we find that $\hbxp,\hat \bup$ scale like $a$, while  $\hat z,\hat u^z$  scale like $a^2$;
hence the trajectory goes to a straight line in the limit $a\!\to\!\infty$. 
This is due to  magnetic force $q\bb \wedge \bB$ in (\ref{EOM}).

\item  {\bf Corollary} \
The final 4-velocity $u$ and energy gain  read
\be
\bup_f\!=\!\hbup(\infty), \quad u^z_f=\E_f= \frac 12\bup_f{}^2=\gamma_f\!-\! 1;
 \label{Lawson-1} 
\ee
Both are very small if the pulse modulation $\epsilon$ is slow [extremely small if 
\ $\epsilon\!\in\!{\cal S}(\mathbb{R})$ \ or \ $\epsilon\!\in\!C^\infty_c(\mathbb{R})$].
\end{itemize}

\noindent
We compare this result with the socalled {\it Lawson-Woodward}  or  {\it General Acceleration} 
Theorem \cite{Law84,Pal88,Pal95,EsaSprKra95}.
This   states that, in spite of large energy variations during the interaction, 
the final energy gain $\E_f$ of a charged particle ${\cal P}$ interacting
with an EM field is zero if: 

\begin{enumerate}

\item the interaction occurs in $\b{R}^3$ vacuum (no boundaries); 

\item   $\bE_s=\bB_s=\0$  and $\Bep$ is very slowly modulated;  

\item   $v^z\simeq c$ along the whole acceleration path; 

\item  nonlinear  (in $\Bep$) effects $q\bb\!\wedge\! \bB$ are negligible; 

\item  the power radiated by ${\cal P}$ is negligible. 

\end{enumerate}

\noindent 
Our Corollary, as Ref. \cite{TrohaEtAl99}, states that the final energy gain 
is zero also if we relax iii), iv), but the EM field is a {\it plane} travelling wave. 
To obtain a non-zero $\E_f$ one has to violate some other  conditions 
of the theorem, as e.g. we see in next cases.

\subsubsection{Case \ $\bE_s=0$,  $\bB_s=B^z_s\bk $, and cyclotron autoresonance.}

\begin{figure}[ht]
\includegraphics[width=7.8cm]{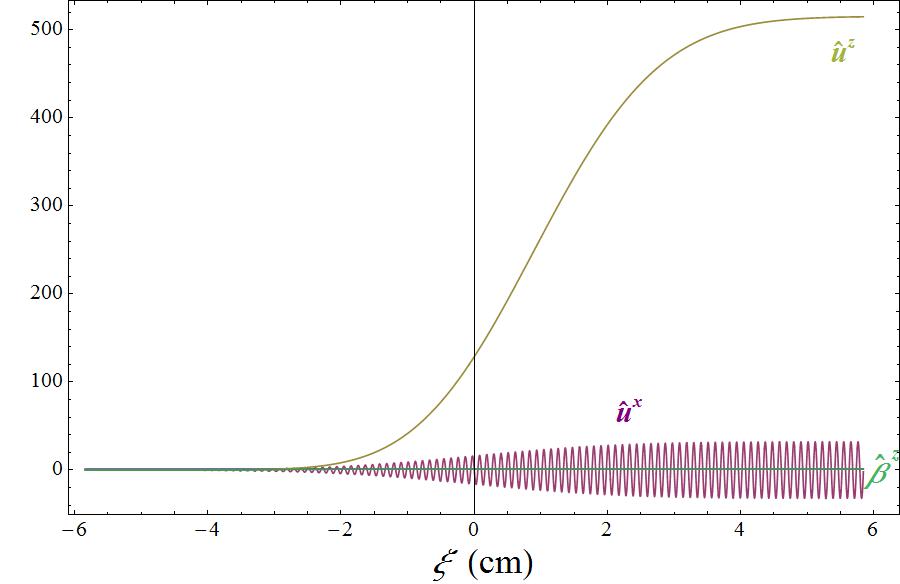} \hfill \includegraphics[width=7.8cm]{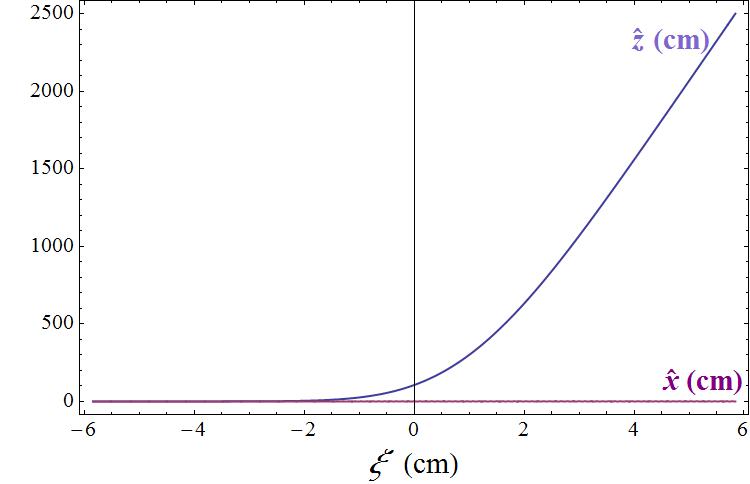}\\[12pt]
\includegraphics[width=16cm]{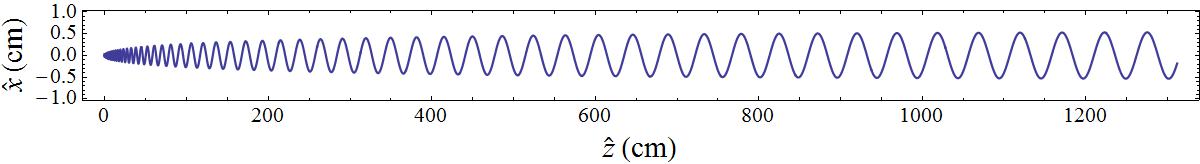}
\caption{The electron 
motion  (\ref{SolEqBz}) (up) and the $zx$-projection of the corresponding trajectory (down) induced 
in a longitudinal magnetic field $B^z\!=\! 10^5$G by a circularly polarized modulated EM wave  (\ref{modulate}) with wavelength $\lambda\!\equiv\!2\pi/k\!=\!1$mm, 
$b \!=\! k=\!58.6$cm$^{-1}$, gaussian enveloping amplitude $\epsilon(\xi)=\EEpM\exp[-\xi^2/2\sigma]$ with
$ \sigma \!=\! 3$cm$^2$ 
and \ $a_0\!\equiv\! e \EEpM/kmc^2\!=\!0.15$,  trivial initial conditions ($\bx_0 \!=\!  \bu_0 \!=\! 0$),  giving $\E_f\!\simeq\! 28.5$.}
\label{Bz=const}      
\end{figure}
Eq. (\ref{SolEqBzEzg}) becomes \ $\hat s\equiv 1$ and, since
$(1\!-\!\kappa\xi)^{ib/\kappa}=\exp\left[\frac{ib}{\kappa}\log(1\!-\!\kappa\xi)\right]=\exp\left[-\frac{ib}{\kappa}\kappa\xi+O(k)\right]$
reduces to $e^{-ib\xi}$ as $\kappa\to 0$, 
\bea
\ba{l}
\displaystyle  ( \hat x \!+\! i \hat y)(\xi)\!=\!\!\int^\xi_0\!\!\!\!  d\zeta \, e^{ib(\zeta\!-\!\xi)}(w^x\!\!+\! iw^y) (\!\zeta\!),
 \qquad \hbup\!=\!\hbxp{}'\!,\\[12pt]
\displaystyle  \hat u^z\!=\!\hat z'\!=\!\frac {\hbup{}^2}{2}\!=\!\E\!=\!\hat\gamma\!-\!1,  \qquad 
\hat z(\xi)\!=\!\!\int^\xi_0\!\!\!\!  d\zeta \:\frac {\hbup{}^2\!(\zeta)}{2}.
\ea   \label{SolEqBz}
\eea
The solution (\ref{SolEqBz}) reduces to that of \cite{KolLeb63,Dav63} 
if $\Bep$  is monochromatic. This leads to {\it cyclotron autoresonance} if 
$-b\!=\!k\!=\!\frac{2\pi}{\lambda}\!\gg\!\frac 1l$. In fact, this conditions ensures that
the cyclotron frequency associated to $B^z_s$ equals the EM wave frequency,
implying an accelerated transverse motion as in a cyclotron (with a spiral trajectory); since also the transverse
magnetic field oscillates with the same frequency, the associated (longitudinal) magnetic force keeps always the same sign and thus accelerates the particle in the $z$-direction
during the whole EM pulse.
If for simplicity  $\Bep$ is slowly modulated, circularly polarized then \ $w^x(\xi)\!+\!iw^y(\xi)\!\simeq\!  e^{ik\xi} {\rm w}(\xi)$,  
where ${\rm w}(\xi)\! \equiv\! q\epsilon(\xi)/kmc^2$, and 
\bea
( \hat x \!+\! i \hat y)(\xi)\simeq  i  W\!(\xi)e^{ik\xi} , 
 \qquad (\hat u^x\!+\! i \hat u^y)(\xi)
\simeq -ke^{ik\xi}W(\xi),     \nonumber                      \label{ApprSolEqBz-2}
\eea
where \ 
$W(\xi) \equiv\! \int^\xi_0\!\!  d\zeta \, {\rm w}(\zeta)\!>\!0$ \ grows with $\xi$. 
In particular if $\Bep(\xi)\!=\!\0$ for $\xi\!\ge\! l\!\equiv$, then for such $\xi$
$$
\hat z'(\xi)\!\simeq\!  \frac {k^2}2 W^2(l)\!\simeq\!2\E_f, \qquad
\frac{|\hbxp{}'(\xi)|}{\hat z'(\xi)}\!\simeq\!\frac 2 {k W\!(l)}\!\ll\! 1; 
$$
the final energy gain and collimation are noteworthy  resp.
by the first, second relation.

\subsubsection{Case \ $\bE_s=E_s^z\bk$, \ $\bB_s=\0$}
\label{Ezconst}

Then the solution (\ref{SolEqBzEzg}) reduces to\footnote{Clearly the domain of definition is $\xi\in[0,\xi_f[$, where $\xi_f\le 1/\kappa$; however $\hat t(\xi_f)=\infty$, so that as a function of $t$ 
the solution is defined in all of $[0,\infty[$, as required  \cite{FioJPA18}.}  \ $\hat s(\xi)=1\!-\! \kappa\,\xi $,
\be
 ( \hat x +i \hat y)(\xi)=\!\!\int\limits^\xi_0\!\!\!  dy \,
 \frac{(w^x\!\!+\!iw^y) (y)} {1\!-\!\kappa y},\quad
\hat z(\xi)=\!\!
\int\limits^\xi_{0}\!\!\!\frac {dy}2 \left\{\!\frac {1\!+\! \hat v(y)}{[1\!-\!\kappa y]^2}\!-\!1\right\}\!;
\label{Ezcost}
\ee
if $\Bep$ is slowly modulated then by (\ref{DeltaHf})  the final energy gain \
$\E_f\simeq \int^{l}_{0}\! d\xi \: \frac {-\kappa \hat  v(\xi)}{2\hat s^2(\xi)}$ 
\ is negative if  $\kappa\!>\!0$, positive if   $\kappa\!\le\!0$
and has a unique maximum point $\kappa_{\scriptscriptstyle M}<0$ if $\epsilon(\xi)$ has a finite support with
a unique maximum. An acceleration device based on this solution could be as follows: 
at $t=0$ the particle  lies at rest with $z_0\!\lesssim\! 0$, just at the left of a metallic grating $G$ contained in the
 $z\!=\!0$ plane and set at zero electric  potential;
another metallic plate $P$  contained in a  plane $z\!=\!z_p\!>\!0$  is set at electric potential $V=V_p$.
A short laser pulse $\Bep$ hitting the particle boosts it beyond $G$
through the ponderomotive force; choosing $qV_p\!>\!0$ implies $\kappa\!=\!-qV_p/z_p mc^2\!<\!0$, and a 
backward longitudinal electric force $qE^z_s$. If  $qV_p$ is large enough, then $z(t)$ reaches a maximum
smaller than $z_p$, then is accelerated backwards and exits the grating
with energy $\E_f$ and negligible transverse momentum. 
However, a large $\E_f$ requires extremely large $|V_p|$, far beyond the material breakdown threshold,
what prevents its realization as a static field (namely, sparks between $G,P$ would arise and rapidly reduce $|V_p|$).
Alternatively, one can make the pulse itself generate such large  $|E_s^z|$ within a plasma at the right time, so as to induce
the {\it slingshot effect} \cite{FioFedDeA14,FioDeN16,FioDeN16b}; this is briefly explained at the end of next section.

\section{Impact of a short laser pulse onto a cold diluted plasma}
\label{Plasmas}

\begin{figure}[h]
\includegraphics[width=7.8cm]{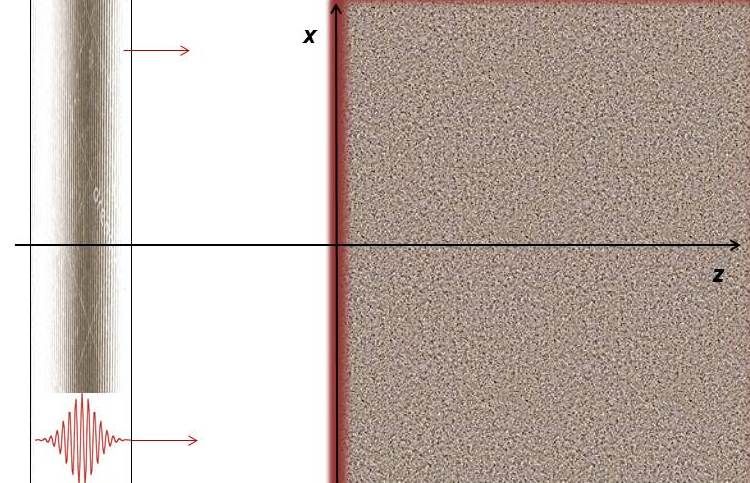}\hfill\includegraphics[width=7.8cm]{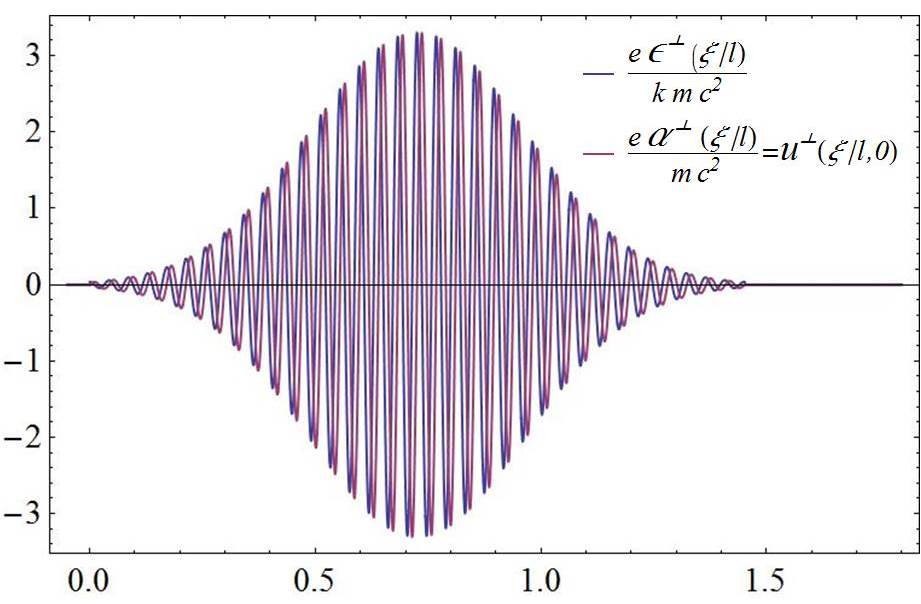} 

\caption{Left: \ a plane EM wave of finite length approaching normally a plasma in equilibrium. \
Right:  \  normalized EM wave $\Bep$ (blue)  with carrier wavelength $\lambda\!=\!0.8\mu$m, linear polarization, 
gaussian modulation $\epsilon(\xi)=\EEpM\exp[-\xi^2/2\sigma]$ with
$ \sigma \!=\! 20\mu$m$^2$, \ $a_0\!\equiv\!e \EEpM/kmc^2\!=\!3.3$ (whence the average pulse intensity is
$10^{19}$W/cm$^2$); the associated $\bup_e$ is painted purple.  $l$ is the length of the $z$-interval  where the amplitude $\epsilon$ overcomes all ionization thresholds  of the atoms of the  gas  yielding the plasma;
here we have chosen helium, whence  $l\!\simeq\! 40\lambda  \!=\!32\mu$m, and the thresholds for $1^{st}$ and $2^{nd}$ ionization  are overcome  almost simultaneously.
}
\label{Plane}
\end{figure}

\subsection{Our simplified hydrodynamic model and its range of validity}

We distinguish the types of particles (electrons and ions, possibly of various kinds)   composing the plasma  by an index $h$, and denote by $q_h,m_h$  their charges, masses.
Assume that the plasma is initially in hydrodynamic conditions with all initial  data  not depending on $\bxp$; these data consist of EM fields of the form (\ref{EBfields}) as well as the set of initial values of the
Eulerian velocity $\bv_h$ and density $n_h$ of each fluid consisting of the same
type of particles. Consequently also the solutions of
the Lorentz-Maxwell  and continuity equations for  $\bB,\bE,\bu_h,n_h$  do not depend on $\bxp$, nor do the 
displacements $\Delta\bx_h\equiv \bx_h(t,\bX)\!-\!\bX$ on $\bXp$. \ Here $\bx_h( t,\!\bX)$ is the position at time $t$
of the material element of the  $h$-th  fluid  with initial position $\bX\!\equiv\!(X,\!Y,\!Z)$; $\bX_h( t,\bx)$
is the inverse of $\bx_h( t,\!\bX)$ (at fixed $t$); $\bb_h\!=\!\bv_h/c$, etc.
For brevity   we refer to the particles (of the $h$-th type) contained: in this material element, as the  `$\bX$ particles' (e.g. `$\bX$ electrons');  in a 
region $\Omega$, as the `$\Omega$ particles'; in the layer between $Z,Z\!+\!dZ$,
as  the `$Z$ particles'.
More specifically, we consider  (fig. \ref{Plane}) a very short and intense EM plane wave 
(laser pulse) 
(\ref{aa'}a) hitting normally a cold  plasma  initially in equilibrium, possibly
immersed in a static and uniform magnetic field $\bB_s$ (actually, the plasma may be locally  obtained from a gas ionized by the very high electric field of the pulse itself). 
The initial conditions are:
\bea 
\ba{l}
n_h(0,\bx)\!=\!0\quad \mbox{if }\: z\!\le\! 0,\qquad \bu_h(0,\bx)\!=\!\0,\qquad
 j^0(0,\bx)\!=\!\sum\limits_h\!q_hn_h(0,\bx)  \!\equiv\! 0, \\[8pt]
\bE(0 ,\bx)\!=\!\Bep(-z),\qquad\quad \bB(0,\bx)\!=\!\bk\wedge\Bep(-z) +\bB_s,
\ea                                                       \label{incond0}
\eea
whence the 4-current density $j\!=\!(j^0,\bj)\!=\!\sum_hq_h n_h(1,\bb_h)$
is zero at $t=0$.
Then the Maxwell equations \ $\nabla\!\cdot\!\bE\!=\!4\pi j^0$,
$ \partial_t E^z\!/c +\!4\pi j^z \!\!=\!(\!\nabla\!\!\wedge\!\bB)^z\!\!=\!0$ imply  \cite{Fio14JPA}
\be
E^{{\scriptscriptstyle z}}(t,\!z)=4\pi \sum_h q_h
\widetilde{N}_h[ Z_h(t ,z)], \qquad \widetilde{N}_h(Z)\!:=\!\int^{Z}_0\!\!d \zeta\,n_h(0,\!\zeta);
         \label{expl}
\ee
using (\ref{expl}) to express $E^z$ in terms of the (still unknown) longitudinal motion 
[$ Z_h(t ,\cdot)$  is the inverse of $z_h(t,\cdot)$] we  reduce the number of unknowns by one. 

Define $\Bap$ as in (\ref{defBap}). 
\ In the Landau gauges  (\ref{incond0}) are compatible 
with the following initial conditions for the gauge potential: 
\bea
 \bA(0 ,\bx)=\Bap(-z)\!+\!\bB_s\!\wedge\!\bx/2,  
\qquad \partial_t \bA(0, \bx)=-c\Bep(-z),
\label{asyc'}
\eea 
$A^0(0,\!\bx)=\partial_t A^0(0,\!\bx)=0$. \ Eq (\ref{incond0}-\ref{asyc'})  and causality imply that  \ $\bx_h(t,\bX)\!=\!\bX$, 
$\bAp(t ,\bx)\!\equiv\! \bB_s\!\wedge\!\bx/2$ if  $ct\!\le\! z$,    $\bj\!\equiv\!\0$ if  $ct\!\le\! |z|$.  
$\bAp$ is coupled to the current through $\Box\bAp=4\pi\bjp$. 
Equipped with (\ref{asyc'}) the latter amounts to the integral equation
\be
\bA\!^{{\scriptscriptstyle\perp}}\!\!-\!
\Ba\!^{{\scriptscriptstyle\perp}}\!-\!\frac 12\left(\bB_s\!\wedge\!\bx\right)^\perp \!=\!
2\pi \!\! \int \!\!\!  d\!s   d\zeta\, \theta(ct\!\!-\!\!s\!-\!|z\!\!-\!\!\zeta|)
\theta(\!s\!) \,\bjp\!\left(\frac sc,\zeta\right);  
       \label{inteq1}
\ee
here we have used  the Green function of the d'Alembertian $\partial_t^2/c^2\!-\!\partial_z^2$ \ in dimension 2 ($\theta$ stands for  the Heaviside step function). 
The right-hand side (rhs) is zero for $t\le 0$ ($t=0$ is the beginning  of the laser-plasma interaction). 
Within {\it short} time intervals $[0,t']$ (to be determined {\it a posteriori})  we can thus:
approximate \ $\bAp(t,z)\simeq \Bap(ct\!-\!z)\!+\!\left(\frac {\bB_s}2\!\wedge\!\bx\right)^\perp$; \ 
also neglect the motion of  ions with respect to the motion  of the (much lighter) electrons. Hence we set \ $z_p(t,Z)\!\equiv\!Z$,
\ and the proton density $n_p$  (due to ions of all kinds) equals the initial one and therefore  the initial  electron density $\widetilde{n_0}(z)\!:=\!n_e(0,z)$, by the initial electric neutrality of the plasma. 

We now adopt $\xi$ instead of $t$ as the independent variable.
Setting $\widetilde{N}(Z)\!:=\!\int^{Z}_0\!\!d \zeta\,\widetilde{n_0}(\zeta)$,
the   equations (\ref{equps0}) \& initial conditions for the electron fluid amount to
\be
\ba{l}
m c^2\hat s_e'(\xi,Z)=4\pi e^2 \left[
\widetilde{N}( \hat z_e ) \!-\! \widetilde{N}( Z)\right] + e(\Delta\hbxp_e{}'\!\wedge\!\hbBp_s)^z\!, \\[10pt]
m c^2 \hbup_e{}'(\xi,Z) =e\Bap{}' - e(\Delta\hat\bx'_e\!\wedge\!\hat\bB_s)^{\scriptscriptstyle \perp}\!,\qquad
\Delta\hat\bx'_e=\frac{\hbu_e(\xi,Z)}{\hat s_e(\xi,Z)}
\ea      \label{equps1}
\ee
\be
\Delta\hat\bx_e(0,\!\bX)\!=\!0, \qquad \hat\bu_e(0,\!\bX)\!=\!\0\qquad\Rightarrow \quad\hat s_e(0,\!\bX)\!=\!1.     
\label{incond}
\ee
\smallskip
This makes eq. (\ref{equps1}) a family parametrized by $Z$ of {\it decoupled ODEs} 
  (of the type considered in section \ref{GenRes}) in the unknowns \ $\Delta\hat\bx_e,\hat s_e$, $\hbup_e$, rather than a set of PDEs. 
After solving these equations and inverting $\hat z_e(x,Z)$,  all Eulerian fields $f(ct,z)$ 
will be obtained from
$\hat f(\xi,Z)$ by the replacement $(\xi,Z)\mapsto\left(ct\!-\!z,\hat Z_e(ct\!-\!z,z)\right)$.

\noindent 
If $\bB_s\!=\!\0$, then as in section \ref{A=A(t,z)}  (\ref{equps1}b)  is solved by $\hbup_e(\xi,Z)\!=\! e\Bap\!(\xi)/mc^2$,  
while, abbreviating $\hat\Delta\!\equiv\!\Delta\hat z_e$ and 
$ v\!\equiv\!\hat\bu^{{\scriptscriptstyle\perp}2}$,    (\ref{equps1}a)  and the
$z$-component of (\ref{equps1}c) take   \cite{FioDeN16,FioDeN16b} the form of (\ref{heq1r}),
\bea
\hat\Delta'\!=\displaystyle\frac {1\!+\! v}{2\hat s^2}\!-\!\frac 12, \qquad
\hat s'_e\!=\frac{4\pi e^2}{mc^2} \left\{ 
\widetilde{N}[\hat z_e] \!-\! \widetilde{N}(Z) \right\}.  \label{heq1} 
\eea
For every $Z$ (\ref{heq1}) have the form of Hamilton equations \ $q'=\partial \hat H/\partial p$, $p'=-\partial \hat H/\partial q$  of a 1-dim system: \  $\xi,\hat\Delta, -\hat s_e$  play the role of $t,q,p$, while the Hamiltonian is {\it rational} in $\hat s_e$ and reads
\bea
\ba{l}
\displaystyle\hat H( \hat \Delta, \hat s_e,\xi;Z)\equiv  \frac{\hat s_e^2+1\!+\!v(\xi)}{2\hat s_e}
+ \U( \hat \Delta;Z),  \\[8pt]
\U( \Delta;Z)\!\equiv\!\frac{4\pi e^2}{mc^2}\left[
\widetilde{{\cal N}}\!\left(Z \!+\!  \Delta\right) \!-\!\widetilde{{\cal N}}\!(Z)\!-\! \widetilde{N}\!(Z)   \Delta\right],  \\[8pt] 
\displaystyle\widetilde{{\cal N}}(Z)\equiv
\int^Z_0\!\!\!d\zeta\,\widetilde{N}(\zeta)\!=\!\int^{Z}_0\!\!\!d\zeta\,\widetilde{n_0}(\zeta)\, (Z\!-\!\zeta);
\ea                                \label{hamiltonian}
\eea
we have defined the potential energy $\U$ fixing the free additive constant so that  $\U(0,Z)\!\equiv\! 0$. 
Eq.s (\ref{heq1}-\ref{incond}) can be solved numerically, or by quadrature where
$\Bep(\xi)\!=\! 0$ (i.e. after the pulse). Finally, the equation \ $\hat\bx_e'\!=\!
\hat\bu_e/\hat s_e$ with initial conditions (\ref{incond})  is solved by
\vskip-.2cm
\be
\hat\bx^{\scriptscriptstyle \perp}_e(\xi,\bX)-\bXp=\!\int^\xi_0\!\!\! d\eta \,\frac{\hat\bu^{\scriptscriptstyle \perp}_e(\eta)}{\hat s(\eta,Z)},\qquad
\hat z_e(\xi,Z)-Z=\hat\Delta(\xi,Z).         \label{hatsol'}
\ee
By derivation we obtain several useful relations, in particular
\vskip-.2cm
\be
\frac{\partial Z_e}{\partial z}(t,z)
=\left.\frac{\hat\gamma_e}{\hat s_e\, \partial_Z\hat z_e}\right\vert_{(\xi,Z)=\big(ct\!-z,Z_e(t,z)\big)} .\label{invZtoz}
\ee
\vskip-.1cm
\noindent
Hence the maps \ $\hat \bx_e(\xi,\cdot)\!:\!\bX\!\mapsto\!  \bx$, \
  $\bx_e(t,\cdot)\!:\!\bX\!\mapsto\!  \bx$ \ are invertible
as long as  $\partial_Z\hat z_e\!\equiv\!\partial\hat z_e/\partial Z$
  is positive. 
The approximation on $\bAp\!(t,\!z)$ is acceptable as long as the so determined motion makes   $|\mbox{rhs}(\ref{inteq1})| \!\ll\! |\Bap\! + \frac {\bB_s}2\! \wedge \bx|$;  otherwise rhs(\ref{inteq1}) determines the first correction to $\bAp$; and so on.

Summarizing, the present hydrodynamic model is justified   in a sufficiently short time
interval $[0,t']$
where \ $\partial_Z\hat z_e\!>\!0$, \ $|\mbox{rhs}(\ref{inteq1})| \!\ll\! |\Bap\! + \frac {\bB_s}2\! \wedge \bx|$, \ and the motion of ions is negligible.

\subsection{Some general features of  the  motions ruled by (\ref{heq1}-\ref{incond})}

$\widetilde{N}(Z)$ grows with $Z$, and so does the rhs(\ref{heq1}b) with $\hat\Delta$.
As soon as $v(\xi )$ \  becomes positive  for $\xi\!>\!  0$,  then so do also
$\hat\Delta$ and $\hat s_e\!-\!1$: by (\ref{hatsol'}),  all electrons reached by the pulse start to oscillate transversely and drift forward (pushed by the ponderomotive force); 
the $Z\!\simeq\!0$ electrons leave behind themselves
a  layer of ions $L_t$ of finite thickness $\zeta(t)\!=\!\Delta(t,0)\!=\!\hat\Delta[\tilde\xi(t,0),0]$ 
completely deprived of electrons, see fig. \ref{DensityPlot-Worldlines}.
   $ \hat s_e$ keeps growing  as long as $\hat\Delta\!\ge\!0$, making the 
rhs(\ref{heq1}a)  vanish at the first $\bar\xi(Z)\!>\!0$ where $\hat s_e^2(\bar\xi,\!Z)\!=\!1\!+\!v(\bar\xi)$
and become negative for  $\xi\!>\!\bar\xi$. Hence $\hat\Delta(\xi,\!Z)$ reaches a 
positive maximum 
at $\xi\!=\!\bar\xi(Z)$ and then starts decreasing 
towards negative values (electrons are attracted back by ions in $L_t$).  For $\xi\!>\!l$ energy is conserved,  
the paths in $\big(\Delta,  s_e\big)$ phase space are the level curves (parametrized by $Z$) fulfilling $\hat H(  \Delta,  s_e;Z)\!=\!h(Z)\!=\!
1\!+\!\!\int^l_0 \!d\xi\: v'(\xi)/\hat s(\xi,Z)$ [by (\ref{DeltaHf})]; the dependence 
of $\hat P\!\equiv\!\big(\hat \Delta, \hat s_e\big)$ on $\xi$ is obtained by quadrature.
Since $\U(\Delta;0)\!=\!0$ for $\Delta\!\le\!0$, then $\hat\Delta(\xi,0)\!\to\!-\!\infty$ as $\xi\!\to\!\infty$: 
the $Z\!=\!0$ electrons escape to $z_e\!=\!-\infty$ infinity.  Whereas if $Z\!>\!0$ then $\U(\Delta;\!Z)\!\to\!\infty$ as 
$|\Delta|\!\to\!\infty$, the path is a cycle, and $P(\xi;Z)$ is periodic in $\xi$, with period
\bea
 c\, T(Z) \equiv \Xi(Z)=\!\int
^{\Delta_{{\scriptscriptstyle M}}}_{\Delta_m}\!\! \frac{2\:  d\Delta}
{\sqrt{ 1-\mu^2/[h(Z)\!-\!\U( \Delta;Z)]^2}},  \qquad     \mu^2\!\equiv\!1\!+\!v(l):                 \label{period}
\eea
all $\hat \Delta(\cdot,Z)$ oscillate around zero, i.e. all $Z\!>\!0$ electrons do $T(Z)$-periodic longitudinal oscillations about their initial positions $z=Z$. 

\begin{figure}[t]
\includegraphics[width=7.8cm]{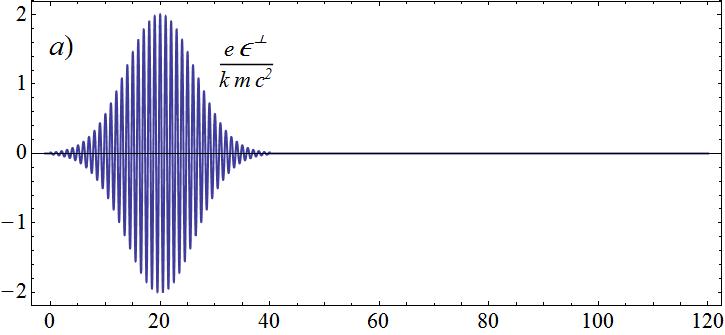}\hfill \includegraphics[width=8cm]{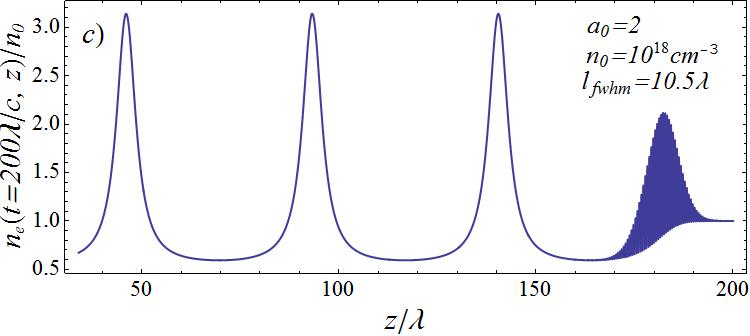}\\
\includegraphics[width=7.8cm]{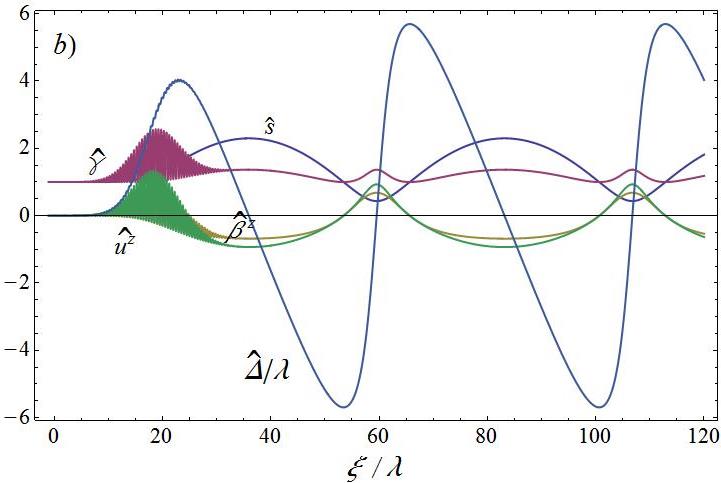} \hfill
\includegraphics[width=8cm]{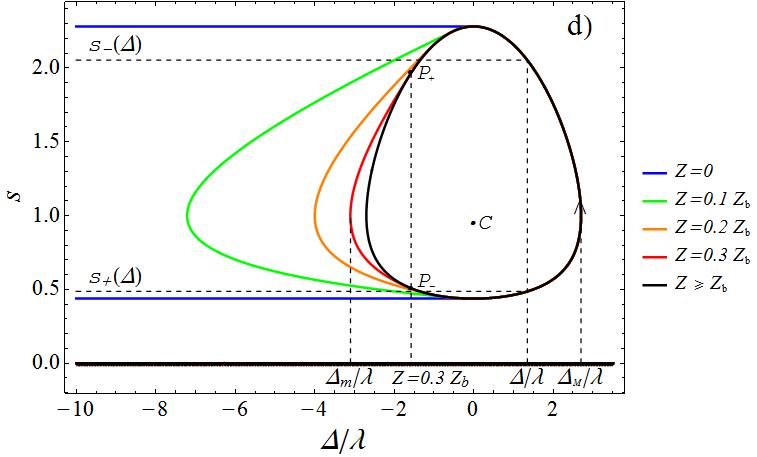}
\caption{
\ a)  Normalized gaussian EM pulse of {\it full width at half maximum} $l_{fwhm}\!=\!10.5\lambda$, linear polarization, 
peak  amplitude $a_0\!\equiv\!2\pi eE^{\scriptscriptstyle \perp}_{\scriptscriptstyle M}/mc^2k\!=\!2$
(leading to a peak  intensity $I\!=\!1.7\!\times\!10^{19}$W/cm$^2$ if $\lambda=0.8\mu$m). \ \ b) \ Corresponding  
solution of (\ref{e1}-\ref{e2}) for $Z\!\ge\!Z_d$, with
$\widetilde{n_0}(Z)\!=\!n_0\theta(Z)$, $n_0\!\equiv\! 10^{18}$cm$^{-3}$
  (whence $h\!=\!1.36$); as anticipated, $\hat s$ is indeed  insensitive to fast oscillations of $\Bep$. \
c) \ Corresponding normalized electron density inside the bulk as a function of $z$ at \ $ct=200\lambda$. \ \ 
d) \ Phase portraits for  the same $\widetilde{n_0}(Z)$, $h\!=\!1.36$, $\mu=1$. 
}
\label{graphs}
\end{figure}

\smallskip
Assume now that the electron density has an upper bound and
becomes a constant well inside the bulk: 
$0\!<\!\widetilde{n_0}(z)\!\le\! n_b$ if $z\!>\!0$ and 
$\widetilde{n_0}(z)\!=\!n_0$  if $z\!\ge\! z_s$, for some
constant $n_b\!\ge\!n_0$,  $z_s\!\ge\! 0$. Then
there exist $Z_b\!>\!0$ and $Z_d\!>\!Z_b,z_s$ such that: 

\begin{enumerate}
\item the $Z\!\in]0,\!Z_b[$ 
electrons exit and re-enter the bulk; 

\item the $Z\!\ge\!Z_b$ electrons remain inside the bulk,
i.e.  $\hat z_e\!\ge\! 0$ for all $\xi$;

\item the $Z\!\ge\!Z_d$ electrons fulfill
$\hat z_e\!\ge\! z_s$  for all $\xi$. 
\end{enumerate}
For the latter $\widetilde{n_{0}}(\!\hat z_e\!)\!\equiv\!n_0$,  $\U(\Delta,Z)\!\equiv\!M\Delta^2\!/2$, so that (\ref{heq1}-\ref{incond}) no longer depends on $Z$ and reduces to the {\it same} Cauchy problem  {\it for all $Z\ge Z_d$}:
\bea
&& \Delta '=\displaystyle\frac {1+v}{2s^2}\!-\!\frac 12,\qquad\quad
s'=M\Delta,\qquad M  \!:=\!\frac{4\pi e^2n_0}{mc^2}\!\equiv\!\frac{\omega_p^2}{c^2}, \label{e1} \\[6pt] 
&&  \Delta (0)\!=\!0, \qquad\qquad\qquad   s(0)\!=\! 1.\label{e2}
\eea
Eq. (\ref{e1}) is the equation of motion of a  relativistic harmonic oscillator  (with a forcing term $v$ for $0<\xi< l$); consequently, also the final energy $h$ is the same for all $Z\!\ge\!Z_d$.
We denote as $\Xi_{{\scriptscriptstyle H}}(n_0)\!=\!cT_{{\scriptscriptstyle H}}(n_0)$ the corresponding plasma period (\ref{period}) 
(recall that $T_{{\scriptscriptstyle H}}\!\ge\!T_{{\scriptscriptstyle H}}^{{\scriptscriptstyle nr}}\!\equiv\!2\pi/\omega_{p}
\!=\!\sqrt{\!\pi m\!/n_0e^2}$,  which is the non-relativistic limit of $T_{{\scriptscriptstyle H}}$\footnote{When $\hat v\!=\!0$ then (\ref{e1}) implies $\Delta''\!=\!-M \Delta/\hat s^3$. In the nonrelativistic regime $\hat s\!\simeq\! 1$,  $\tilde \xi(t)\!\simeq\! ct$, $c d/d\xi\!\simeq\! d/dt$,
and this becomes the nonrelativistic harmonic equation $ \ddot \Delta\!=\!- \omega_p^2\Delta$.}). Thus for all $Z>Z_d$  electrons it is $\hat z_e(\xi,Z)=Z\!+\!\Delta(\xi)$ and hence $ z_e(t,Z)=Z\!+\!\Delta(ct\!-\!z)$ for all $\xi$. The
restriction of $z_e(t,\cdot)$ to $[Z_d,\infty[$ is invertible 
[with inverse \ $Z_e(t,z)=z-\Delta(ct\!-\!z)$]: no $Z$-electron layer with $Z\!\ge\!Z_d$ can collide with another one. This justifies there  the hydrodynamical picture used so far.
All Eulerian fields are found to  depend  on $t,z$ only through $ct\!-\!z$, e.g. $\bu(t,z)=\hat\bu(ct\!-\!z)$,...: a plasma travelling-wave with spacial period 
$\Xi_{{\scriptscriptstyle H}}(n_0)$ and phase velocity $c$  trails the pulse for
$z\ge Z_d$ \cite{FioRM19
}.
On the other hand, plasma wave breakings \cite{Daw59}, i.e. collisions  among (sufficiently close) $Z$-electron layers with $Z\!<\!Z_d$, occur only at sufficiently large times,  due to the nontrivial $Z$-dependence of (\ref{period}) in the regions where 
$\widetilde{n_0}(Z)$ is non-homogenous (in particular near the vacuum-plasma interface $Z\!\sim\!0$). 
There our hydrodynamic description is globally self-consistent for $t\!<\! t_c$ ($t_c$ stands for the time of the first wave-breaking) and allows to determine \cite{FioRM19
} $t_c$ and where   wave-breakings occur;
the use of kinetic theory  (a statistical description of the plasma in phase space that takes collisions into account) is necessary  after the first wave-breaking. 
As known, a moderate wave-breaking is actually welcome as a possible injection mechanism in the plasma wave of a bunch of electrons  fast enough and in phase to be accelerated 'surfing' the wave, according to the LWFA mechanism \cite{Tajima-Dawson1979}; a phase of the right sign arises where $\widetilde {n_0}(Z)$ decreases.

\begin{figure}
\includegraphics[width=5.2cm]{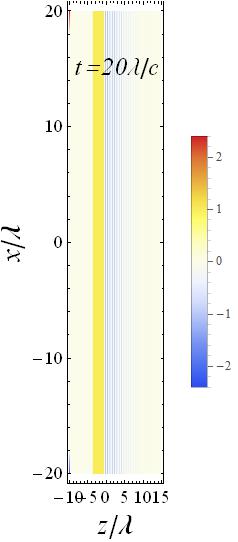}
\hfill
\includegraphics[width=10.6cm]{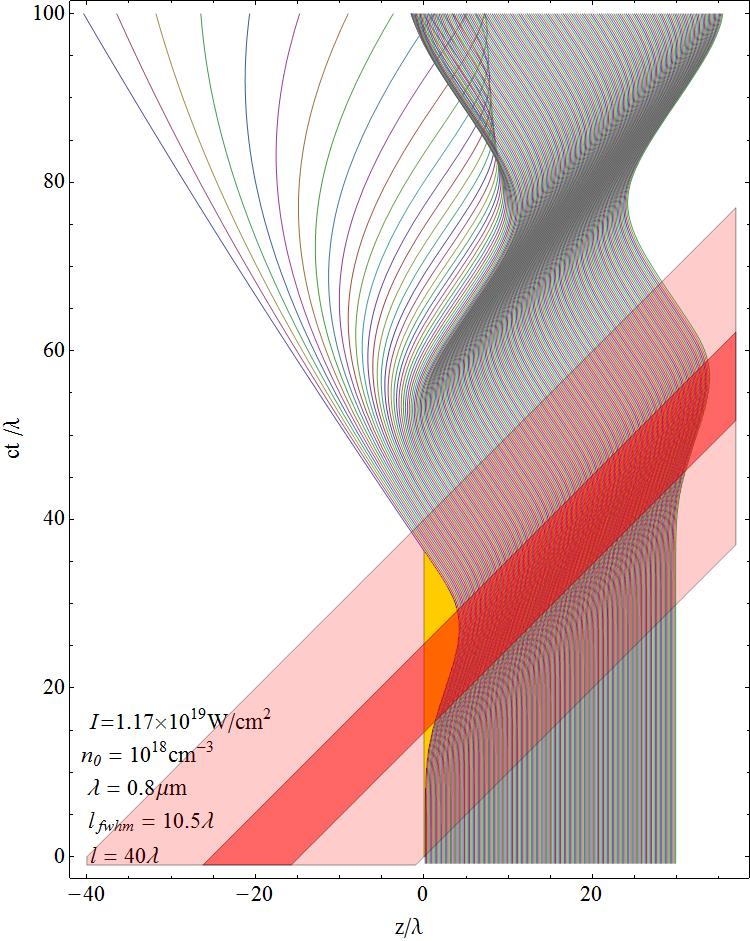}
\caption{Conditions as in fig. \ref{graphs} ($\widetilde{n_0}(Z)\!=\!n_0\theta(Z)$). \ Left: Normalized total charge density $1\!-\!n_e/n_0$ at $ct=20\lambda$; the yellow part denotes the positive ion layer $L_t$,  the thin blue stripes denote the negative parts (due to an excess of electrons). \
Right: The corresponding worldlines of $Z$-electrons; for $Z\ge Z_d\simeq 10\lambda$ they do not intersect, whereas the $Z< Z_d $ ones that first intersect
do  about after $5/4$ oscillations induced by the pulse, here at $t=t_c\!\simeq\! 84\lambda/c\!=\!224$fs. In pink the support $0\le ct\!-\!z\le l$ of the pulse, in  red the 'effective support' $\frac{l\!-\!l_{fwhm}}2\le ct\!-\!z\le \frac{l\!+\!l_{fwhm}}2$ of the pulse,in
yellow the evolution of the pure-ion layer $L_t$.
}
\label{DensityPlot-Worldlines}       
\end{figure}

Let $T_{{\scriptscriptstyle H}}(n_b)$ is the plasma period (\ref{period}) associated to the
electron density $\widetilde{n_0}(Z)\equiv n_b$. If 
\be
2l\lesssim \Xi_{{\scriptscriptstyle H}}(n_b)\equiv c T_{{\scriptscriptstyle H}}(n_b)
                  \label{Lncond}
\ee
(i.e. the pulse is sufficiently short) then $\hat\Delta(l,\!Z)\!\ge\! 0$ for all $Z$:  the pulse overcomes all electrons
before they overshoot their initial $z$-coordinate; in particular, 
it has completely entered 
the bulk before any  small-$Z$ electron gets out of it, namely before 
$L_t$ is refilled. Condition (\ref{Lncond})  also secures  that the spacial period of the plasma wave is larger that the pulse length.

In fig. \ref{graphs} we plot the solution  corresponding to the pulse of fig. \ref{Plane}-right (with $l\!\simeq\! 40\mu$m, $l_{fwhm}\!\simeq\! 10.5\mu$m) and  to \
$n_0\!=\!  10^{18}$cm$^{-3}$; \ $s(\xi)$ is indeed insensitive to the fast oscillations of $\Bep$ (see remark  \ref{A=A(t,z)}.\ref{insensitive}),  $\Delta(\xi)$ grows positive  for small $\xi$. The other unknowns are obtained through (\ref{hatsol}).
After the pulse is passed  the solution becomes periodic with period $\Xi_{\scriptscriptstyle H}\!\simeq\!  47\mu$m. These  $l_{fwhm}$, $n_0$ 
fulfill (\ref{Lncond}).
Replacing these solutions  in the rhs(\ref{inteq1}) we find that $\bAp\!\simeq\!\Bap$ is indeed verified 
at least  for $t\!<\! t_c\!\simeq\!  5 \xi_{\scriptscriptstyle H}/c$. 
In fig. \ref{DensityPlot-Worldlines} we plot the charge density just after the
pulse has hit the plasma and the worldlines of the $Z$-electrons for $0<Z<30\lambda$
and $0\le t\le 100\lambda/c\simeq 267$fs.

\subsection{Comparison with the `comoving frame' approach}

The Lorentz-Maxwell and continuity equations  in a (homogeneous) 
plasma admit \cite{AkhPol56} travelling wave solutions where the 
 EM field (plane wave laser pulse) has the form $\bE(t,\bx)\!=\!\Bep(v_lt\!-\!z)$, $\bB(t,\bx)=\bk\wedge\Bep(v_lt\!-\!z)$; i.e. the pulse travels with a (constant) velocity $\bv_l=v_l\bk$, $v_l\le c$.
If $v_l\!<\!c$, in studying the electrodynamics  (plasma waves, LWFA, ...)
induced by such a pulse 
it is common and useful to adopt the `comoving frame' approach.
This consists in describing the physics not w.r.t. the inertial reference frame $\R$ of the laboratory, but w.r.t. a one $\R'$ comoving with the pulse, i.e. moving
with velocity $\bv_l$ with respect to (w.r.t.) $\R$. Setting $\beta_l=v_l/c$, $\gamma_l^{-1}=\sqrt{1\!-\!\beta_l^2}$, the spacetime coordinates w.r.t. $\R,\R'$ are related
by the Lorentz transformation 
\be
ct'=\gamma_l(ct-\beta_l z), \qquad \bxp{}'=\bxp, \qquad z'= \gamma_l(z-\beta_l ct),
\label{LorentzTransf}
\ee
and more generally the components of all 4-tensors  [in particular the canonically conjugated 4-momentum $(H,c\bP)$ and the EM tensor $(F^{\mu\nu})$] w.r.t. $\R,\R'$ are related by this Lorentz transformation and can be expressed in terms of $\rx'$; the pulse part $\Bep{}'$ of the EM field is a function of $t'$ only.
Therefore adopting $t'$ as the independent variable has the same advantage described before, i.e. $\Bep{}'(t')$ appears in the equations of motion of
all charged particles (including the plasma constituents) as  a  known forcing term;
correspondingly, the known expression $\bAp{}'(\rx')=\Bap{}'(t')$ appears in the Hamiltonian of all charged particles. However, this approach is no more applicable 
when $v=c$, for which the transformation (\ref{LorentzTransf}) is ill-defined.

On the contrary, although  \ $\xi=ct\!-\!\beta_l z\propto t'$ \ can no more be interpreted as the time coordinate w.r.t.
an inertial reference frame, in our approach adopting $\xi$ instead of $t'$ as the independent variable  keeps this advantage and allows
to take the limit $v\to c$. 

\subsection{Finite laser spot radius corrections, slingshot effect, and discussion}
\label{3D-effects}

The above  predictions  are based on idealizing the laser pulse as a plane EM wave. In a more realistic
picture the laser pulse is cylindrically symmetric around the $\vec{z}$-axis and has a {\it finite} spot radius $R$. The first rough correction  to the above predictions is that only
the $Z\!\simeq\!0$ electrons inside a cylinder of radius $R$ are pushed forward and leave behind themselves a  cylinder $c_t$ of ions of the same radius and finite height $\zeta(t)\!=\!\Delta(t,0)\!=\!\hat\Delta[\tilde\xi(t,0),0]$  completely deprived of electrons.
Using causality and heuristic arguments we can compute \cite{FioDeN16}
further (rough) corrections, in particular estimate the  motion of the
lateral electrons (LE), just outside $c_t$, toward the axis of $c_t$, 
attracted by the ions. If $R$ is not too small
the $Z$-electrons  with smallest  $Z$ and closest to  the  axis  succeed
in exiting the bulk before the LE reach the axis and close their way out, and
proceed indefinitely in the negative $z$ direction.
As a result, for carefully tuned $R,\widetilde{n_0}(Z) $  [fulfilling
(\ref{Lncond}), in particular] the impact of a very short and intense  laser pulse 
on the surface of a cold low-density  plasma (or gas,  ionized into a plasma by the pulse itself),
as considered e.g. in  fig. \ref{Plane}-right,  may induce, 
beside  a plasma traveling-wave  trailing the pulse,
also the {\it slingshot effect} \cite{FioDeN16,FioDeN16b,FioFedDeA14},
i.e. the backward acceleration and expulsion from the plasma 
 of some surface electrons (those with smallest $Z$ and closest to  the $\vec{z}$-axis)  with remarkable energy.
For reviews see also \cite{Fio14,Fio16b}. 
On the other hand, if $R$ is smaller the LE may close the rear part
of $c_t$ and make it into a ion {\it bubble} (completely deprived of electrons), before any electron gets out of the bulk. Depending on the conditions, the bubble may then disappear or trail  the pulse; in the latter case the LWFA takes place in the particularly favourable {\it bubble regime}  \cite{RosBreKat91,PukMey2002,KosPukKis2004,MaslovEtAl16}.


\end{document}